\documentclass[twocolumn]{aastex63}

\usepackage{caption}
\usepackage{subcaption}
\usepackage{rotating}
\usepackage{hyperref}

\newcommand{\nii}{[{\sc N\,ii}]}
\newcommand{\sii}{[{\sc S\,ii}]}

\usepackage{rotating}
\received{}
\revised{}
\accepted{}
\submitjournal{ApJ}

\shorttitle{High-Spectral Resolution Observations of the Optical Filamentary Nebula in NGC 1275}
\shortauthors{Vigneron et al.}
\graphicspath{{./}{figures/}}
\begin{document}

\title{High-Spectral Resolution Observations of the Optical Filamentary Nebula in NGC 1275}

\correspondingauthor{Benjamin Vigneron}
\email{benjamin.vigneron@umontreal.ca}

%%% HERE JULIE YOU CAN CHOSE THE ORDER OF CO-AUTHORS

\author[0000-0002-2478-5119]{Benjamin Vigneron}
\affiliation{Département de Physique, Université de Montréal, Succ. Centre-Ville, Montréal, Québec, H3C 3J7, Canada}

\author[0000-0001-7271-7340]{Julie Hlavacek-Larrondo}
\affiliation{Département de Physique, Université de Montréal, Succ. Centre-Ville, Montréal, Québec, H3C 3J7, Canada}
\affiliation{Centre de recherche en astrophysique du Québec (CRAQ)}

\author[0000-0003-2001-1076]{Carter Lee Rhea}
\affiliation{Département de Physique, Université de Montréal, Succ. Centre-Ville, Montréal, Québec, H3C 3J7, Canada}
\affiliation{Centre de recherche en astrophysique du Québec (CRAQ)}

\author[0000-0002-2478-5119]{Marie-Lou Gendron-Marsolais}
\affiliation{Instituto de Astrofísica de Andalucía, IAA-CSIC, Apartado 3004, 18080 Granada, España}

\author[0000-0003-4220-2404]{Jeremy Lim}
\affiliation{Department of Physics, The University of Hong Kong, Pokfulam Road, Hong Kong}

\author{Jake Reinheimer}
\affiliation{Department of Physics, University of North Texas, Denton, TX 76203, USA}

\author[0000-0001-5262-6150]{Yuan Li}
\affiliation{Department of Physics, University of North Texas, Denton, TX 76203, USA}

\author[0000-0003-1278-2591]{Laurent Drissen}
\affiliation{Département de physique, de génie physique et d'optique, Université Laval, Québec (QC), G1V 0A6, Canada}
\affiliation{Centre de recherche en astrophysique du Québec (CRAQ)}
\affiliation{Department of Physics and Astronomy, University of Hawai'i at Hilo, Hilo, HI 96720, USA}
\affiliation{Canada-France-Hawaii Telescope, 65-1238 Mamalahoa Hwy, Kamuela, Hawaii 96743, USA}

%À PARITR DE THOMAS (ordre alphabétique pour tout le monde en dessous - alphabétique avec nom de famille). 

\author[0000-0003-2630-9228]{Greg L. Bryan}
\affiliation{Department of Astronomy, Columbia University, 550 West 120th Street, New York, NY 10027, USA}
\affiliation{Center for Computational Astrophysics, Flatiron Institute, 162 5th Avenue, New York, NY 10010, USA}

\author[0000-0002-2808-0853]{Megan Donahue}
\affiliation{Department of Physics and Astronomy, Michigan State University, East Lansing, MI 48824, USA}

\author[0000-0002-3398-6916]{Alastair Edge}
\affiliation{Department of Physics, Durham University, South Road, Durham DH1 3LE, UK}

\author[0000-0002-9378-4072]{Andrew Fabian}
\affiliation{Institute of Astronomy, University of Cambridge, Madingley Road, Cambridge CB3 0HA, UK}

\author[0000-0003-1932-0162]{Stephen Hamer}
\affiliation{Institute of Astronomy, University of Cambridge, Madingley Road, Cambridge CB3 0HA, UK}

\author[0000-0002-3074-9608]{Thomas Martin}
\affiliation{Département de physique, de génie physique et d'optique, Université Laval, Québec (QC), G1V 0A6, Canada}
\affiliation{Centre de recherche en astrophysique du Québec (CRAQ)}

\author[0000-0001-5226-8349]{Michael McDonald}
\affiliation{Kavli Institute for Astrophysics and Space Research, MIT, Cambridge, MA 02139, USA}

\author{Brian McNamara}
\affiliation{Department of Physics and Astronomy, University of Waterloo, 200 University Avenue West, Waterloo, ON N2L 3G1, Canada}

\author[0000-0001-7597-270X]{Annabelle Richard-Lafferrière}
\affiliation{Institute of Astronomy, University of Cambridge, Madingley Road, Cambridge CB3 0HA, UK}

\author[0000-0002-5136-6673]{Laurie Rousseau-Nepton}
\affiliation{Canada-France-Hawaii Telescope, 65-1238 Mamalahoa Hwy, Kamuela, Hawaii 96743, USA}

\author[0000-0002-3514-0383]{G. Mark Voit}
\affiliation{Department of Physics and Astronomy, Michigan State University, East Lansing, MI 48824, USA}

\author[0000-0002-0104-9653]{Tracy Webb}
\affiliation{Department of Physics, McGill Space Institute, McGill University, 3600 rue University, Montreal H3A 2T8, Canada}

\author[0000-0003-0392-0120]{Norbert Werner}
\affiliation{Department of Theoretical Physics and Astrophysics, Faculty of Science, Masaryk University, Brno, Czech Republic}

%% Note that the \and command from previous versions of AASTeX is now
%% depreciated in this version as it is no longer necessary. AASTeX 
%% automatically takes care of all commas and "and"s between authors names.

%% AASTeX 6.3 has the new \collaboration and \nocollaboration commands to
%% provide the collaboration status of a group of authors. These commands 
%% can be used either before or after the list of corresponding authors. The
%% argument for \collaboration is the collaboration identifier. Authors are
%% encouraged to surround collaboration identifiers with ()s. The 
%% \nocollaboration command takes no argument and exists to indicate that
%% the nearby authors are not part of surrounding collaborations.

%% Mark off the abstract in the ``abstract'' environment. 
\begin{abstract}%250 words limit

We present new high-spectral resolution observations (R = $\lambda/\Delta\lambda$ = 7000) of the filamentary nebula surrounding NGC 1275, the central galaxy of the Perseus cluster. These observations have been obtained with SITELLE, an imaging Fourier transform spectrometer installed on the Canada-France-Hawai Telescope (CFHT) with a field of view of $11\text{ arcmin }\times 11 \text{ arcmin}$ encapsulating the entire filamentary structure of ionised gas despite its large size of $80 \text{ kpc}\times50 \text{ kpc}$. Here, we present renewed flux, velocity and velocity dispersion maps that show in great detail the kinematics of the optical nebula at \sii$\lambda6716$, \sii$\lambda6731$, \nii$\lambda6584$, H$\alpha$(6563\AA), and \nii$\lambda6548$. These maps reveal the existence of a bright flattened disk-shaped structure in the core extending to r $\sim 10$ kpc and dominated by a chaotic velocity field. This structure is located in the wake of X-ray cavities and characterised by a high mean velocity dispersion of $134$ km/s. The disk-shaped structure is surrounded by an extended array of filaments spread out to $r\sim 50$ kpc that are 10 times fainter in flux, remarkably quiescent and has a uniform mean velocity dispersion of $44$ km/s. This stability is puzzling given that the cluster core exhibits several energetic phenomena. Based on these results, we argue that there are two mechanisms to form multiphase gas in clusters of galaxies: a first triggered in the wake of X-ray cavities leading to more turbulent multiphase gas and a second, distinct mechanism, that is gentle and leads to large-scale multiphase gas spread throughout the core.

\end{abstract}

%% Keywords should appear after the \end{abstract} command. 
%% See the online documentation for the full list of available subject
%% keywords and the rules for their use.

%% From the front matter, we move on to the body of the paper.
%% Sections are demarcated by \section and \subsection, respectively.
%% Observe the use of the LaTeX \label
%% command after the \subsection to give a symbolic KEY to the
%% subsection for cross-referencing in a \ref command.
%% You can use LaTeX's \ref and \label commands to keep track of
%% cross-references to sections, equations, tables, and figures.
%% That way, if you change the order of any elements, LaTeX will
%% automatically renumber them.
%%
%% We recommend that authors also use the natbib \citep
%% and \citet commands to identify citations.  The citations are
%% tied to the reference list via symbolic KEYs. The KEY corresponds
%% to the KEY in the \bibitem in the reference list below. 

\section{Introduction} \label{sec:intro}

Clusters of galaxies are extended structures hosting several hundred to thousands of gravitationally bound galaxies (e.g. \citealt{bahcall_clusters_1977}, \citealt{abell_catalog_1989}). They are mostly composed of dark matter while galaxies only represent a very small fraction of the cluster's mass (e.g. \citealt{sand_dark_2004}, \citealt{voigt_galaxy_2006}). There is also a third component made of hot X-ray emitting gas at temperatures of $\sim 10^7-10^8$ K that fills the space between the galaxies and is known as the intra-cluster medium or ICM (e.g. \citealt{cavaliere_extragalactic_1971}, \citealt{gursky_detection_1971}, \citealt{hitomi_collaboration_quiescent_2016}).

This ICM accounts for a substantial fraction of the mass of a cluster ($\sim13\%$) and often leads clusters to being classified into two distinct categories depending on their X-ray emission profiles: cool-core clusters with a strongly peaked X-ray emission profile and non-cool core clusters with a more diffuse and uniform X-ray emission profile (\citealt{million2009chandra}, \citealt{hudson2010cool}). The central dominant galaxy located in the core, known as the brightest cluster galaxy (BCG), often exhibits extended filamentary nebulae of optical ionised gas in the case of cool-core clusters (\citealt{crawford_rosat_1999}, \citealt{mcdonald_origin_2010}). These filaments are composed of multiphase gas that have high H$\alpha$ luminosities (e.g. \citealt{conselice2001nature}) and are often co-spatial with cold gas (e.g. \citealt{salome_cold_2006}), as well as soft X-ray gas. The most extended filaments can be mostly devoid of star formation, which excludes photoionization as a primary ionization mechanism (e.g. \citealt{kent_ionization_1979}).

A promising explanation for the formation of the filamentary nebulae surrounding BCGs resides in the precipitation limit hypothesis (e.g. \citealt{gaspari2012cause, gaspari2013chaotic, gaspari2015chaotic}, \citealt{voit2015cooling, voit2015regulation}, \citealt{li2015cooling}, \citealt{voit_global_2017}). Here, the constant cooling by emission of the ICM in cool-core clusters should imply high star formation rates in the cores of clusters. However, since there is limited evidence of correspondingly extremely high star formation inside these galaxies, a mechanism is necessary to stop the gas from cooling and falling in the potential well (e.g. \citealt{crawford_rosat_1999}). In clusters of galaxies, the heating mechanism is thought to be orchestrated by the supermassive black hole residing in the BCG. This mechanism can be seen in action in radio bubbles generated by the central black hole which carve out large X-ray cavities in the ICM, entailing shocks, turbulence, and mixing (e.g. \citealt{graham_weak_2008}, \citealt{randall_very_2015}, \citealt{zhuravleva_turbulent_2014}). This phenomenon will in turn bring enough energy to limit the cooling of the ambient gas and therefore star formation in the galaxy (\citealt{crawford_rosat_1999}, \citealt{best_prevalence_2007}, \citealt{voit_conduction_2008}).               However, the mechanisms explaining how the AGN energy reheats the ICM remains a matter of debate.

The precipitation limit hypothesis proposes that the heating process associated with the activity of the AGN sets in motion flows, which in turn encourage the hot surrounding medium to condense at higher altitudes. Hence, the adiabatic uplift of material through radio bubbles promotes condensation by reducing the ratio of cooling time to free-fall time in some locations. This cycle is analogous to the precipitations occurring on Earth, where raindrops are formed by the uplifting of gas higher in the atmosphere. Moreover, simulations show that the rain of cold gas towards the center of the galaxy will initially feed the central black hole and therefore the power of the flows that it produces, enabling a self-regulated feedback loop (e.g. \citealt{voit_global_2017}). This energy from the black hole will, in turn, heat the ambient medium and increase the ratio of the cooling time to the free fall time so that the precipitation will cease. The filamentary nebulae surrounding BCGs have been interpreted as a tell-tale sign of this model and could therefore offer the possibility to study this hypothesis.

Here, we target the filamentary nebula that surrounds NGC 1275, the BCG located at the center of the Perseus cluster of galaxies. NGC 1275 has been studied extensively at all wavelengths (see \citealt{forman_observations_1972}, \citealt{fabian_wide_2011}, \citealt{salome_very_2011}, \citealt{lim_molecular_2012}, \citealt{nagai_alma_2019}). The cluster hosts several X-ray cavities that originate from multiple generations of radio jets emitted from the active galactic nuclei (AGN) of NGC 1275 (\citealt{1993MNRAS.264L..25B}, \citealt{fabian_chandra_2000}, \citealt{fabian_wide_2011}, \citealt{gendron-marsolais_revealing_2018}). These jets carve the neighbouring ICM, creating buoyantly rising radio-emitting bubbles.

NGC 1275 also displays one of the largest filamentary nebulae known with a size of about $80$ kpc $ \times \text{ } 50$ kpc (e.g. \citealt{mcnamara_optical_1996}, \citealt{conselice2001nature}, \citealt{hamer_optical_2016}). Combined with the proximity of the Perseus cluster, this makes NGC 1275 a target of choice for our understanding of the formation and ionization mechanisms of filamentary nebulae in clusters.

The first observations of the filamentary nebula surrounding NGC 1275 by \cite{minkowski_optical_1957}, \cite{lynds1970improved}, and \cite{rubin1977new} unveiled a high-velocity (HV) feature ($\sim$ 8200 km/s) associated with a forefront spiral galaxy falling onto NGC 1275 (\citealt{2015ApJ...814..101Y}), and a low-velocity (LV) structure ($\sim$ 5200 km/s) linked to NGC 1275. \textit{Hubble Space Telescope} observations of the low-velocity structure then revealed the filamentary appearance of the ionised gas (\citealt{fabian_magnetic_2008}), whereas soft X-ray counterparts were discovered for certain bright filaments with the  \textit{Chandra X-Ray Observatory} (\citealt{fabian_relationship_2003}). Moreover, cold molecular gas has been associated with the emission of the filamentary nebula of NGC 1275 (\citealt{salome_cold_2006}, \citealt{ho_multiple_2009}, \citealt{salome_very_2011}, \citealt{mittal2012herschel}), and also linked to a disk of emission near the galaxy (\citealt{lim_radially_2008}).
% \citealt{nagai_alma_2019}).\\

\begin{figure}
    \centering\includegraphics[width=85mm,scale=0.5]{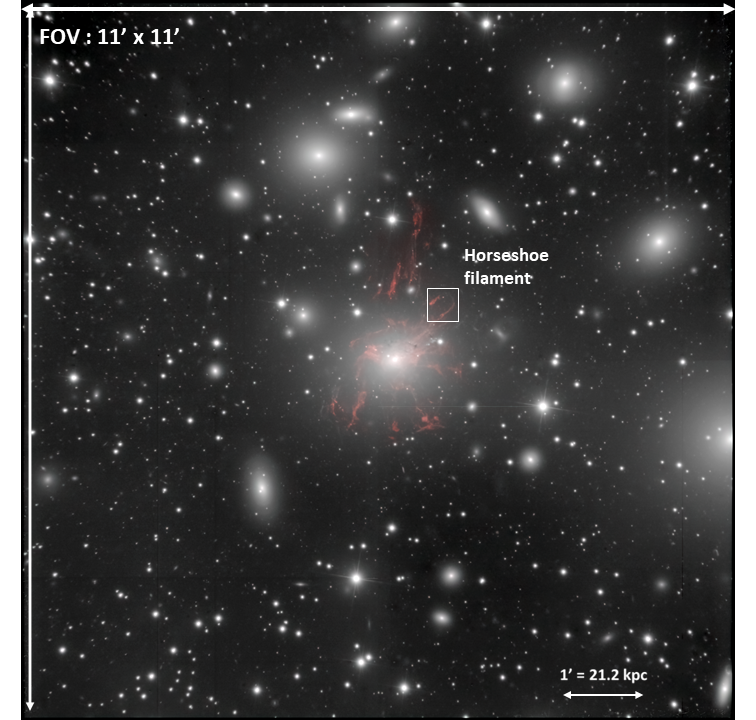}
    \caption{The complete field of view (11' $\times$ 11') of the SITELLE observations of NGC 1275 with full-band SN3 filter (648-685 nm) at high-spectral resolution (R = 7000). Here, we show the integrated flux images in grayscale with the brightest H$\alpha$ filaments in red. The white square denotes the Horseshoe filament.}
    \label{fig:fov}
\end{figure}

Nevertheless, the first detailed observations of the nebula was performed by \cite{conselice2001nature} with high-resolution imaging, integral field, and long-slit spectroscopy (WIYN \& KPNO). These observations brought forth the first velocity map of the central $\sim 45 \text{ arcsec}$ ($\sim 16.5$ kpc) of the nebula. Observations from the Gemini Multi-Object Spectrograph along six slits carefully positioned along certain filaments showed evidence of outflowing gas and flow patterns (\citealt{hatch_origin_2006}). In 2018, the filamentary nebula was then imaged for the first time with SITELLE (\textit{Spectromètre Imageur à Transformée de Fourier pour l’Etude en Long et en Large de raies d’Emission}), an imaging Fourier transform spectrometer (\citealt{2019MNRAS.485.3930D}) installed at the Canada-France-Hawaï Telescope (CFHT) that has an extremely large field of view ($11\text{ arcmin } \times 11\text{ arcmin}$) capable of imaging the nebula in its entirety (see Figure \ref{fig:fov}). \cite{gendron-marsolais_revealing_2018} showed that the velocity structure of the filaments appears to be generally devoid of specific trends or rotation.

%However, these observations converge toward the idea of a formation scenario involving another mechanism for the filamentary nebula surrounding NGC 1275. Indeed, this structure could be formed by the influence of the AGN whose jets create bubbles rising in the ICM (\cite{churazov_evolution_2001}, \cite{reynolds_buoyant_2005}). The north-eastern "Horseshoe" filament of the nebula and its less visible counterpart also support this argument since their shape and symmetric positions are reminiscent of the toroidal flow pattern created by the elevation of a buoyant bubble in a liquid (\cite{fabian_relationship_2003}). Therefore, a formation model through turbulence-induced cooling by radio bubbles can likely be involved.\\

% \textbf{Moreover, clear spatial correlation between several emission structures at different wavelengths has been studied extensively. Indeed, clear detection of molecular gas associated with optical filaments have been reported throughout the nebula and close to the central region, bubbles filled with radio emission have carved the surrounding hot ICM leaving cavities in its X-ray emission. These spatial correlations are of extreme interest to properly understand the complete cycle of feedback and cooling taking place within the center of the galaxy cluster.}\\

In this paper, we present new high-spectral resolution observations of the filamentary nebula surrounding NGC 1275 obtained with SITELLE. The first SITELLE observations presented in \cite{gendron-marsolais_revealing_2018} revealed the kinematics of the ionized nebula, but suffered from poor spectral resolution (R = 1800) that could not spectroscopically resolve the filaments beyond a projected radii of $r\gtrsim 10$ kpc centered on the AGN.

Here, the high-spectral resolution nature of the data at R = 7000 allows us to deepen the scope of the study by performing a detailed analysis of the kinematics of the nebula, in particular the velocity dispersion of the gas. The analysis of the data revealed a central structure displaying higher mean flux and velocity dispersion than the rest of the outer filaments. This disk-shaped structure seems spatially correlated with a similar structure of cold molecular gas, which will be detailed in later sections. This result appears relevant when considering the clear kinematic correlations between the ionized and molecular gas in the central region,
% between the optical data of SITELLE and radio observations of the central structure 
as well as possibly far infrared (IR) [CII] emission lines of the gas surrounding NGC 1275 as observed with Herschel by \cite{mittal2012herschel}.%\textbf{(see Figure \ref{fig:herschel_contours})}.}

In Section 2, we present the new SITELLE observations and the various procedures used during the data analysis. In Section 3 we present and discuss our results. Finally, a summary of our conclusions will be presented in Section 4.

To directly compare our results to those of \cite{gendron-marsolais_revealing_2018} and the Hitomi \cite{hitomi_collaboration_atmospheric_2018}, we also adopt for NGC 1275 a redshift of $z=0.017284$, which implies an angular scale of $21.2 \; \text{kpc arcmin}^{-1}$. This $z$ also corresponds to a luminosity distance of $75.5$ Mpc, assuming $H_0 = 69.6 \; \text{km s}^{-1} \text{Mpc}^{-1}$, $\Omega_M = 0.286$ and $\Omega_{vac} = 0.714$\\

\newpage
\section{Data Analysis} \label{sec:style}

\subsection{Observations with SITELLE}

The filamentary nebula surrounding NGC 1275 was observed in February 2020 with SITELLE during Queued Service Observations 20AD99 by PIs Hlavacek-Larrondo and Rhea. SITELLE is a Fourier transform imaging spectrometer with an incredibly large field of view of 11 arcmins by 11 arcmins and equipped with two E2V detectors of 2048 × 2064 pixels, resulting in a spatial resolution of 0.321 × 0.321 arcsecs. SITELLE was used with the SN3 filter for 4 hours (1710 exposures of 8.42s) including overheads, necessary to obtain a high-spectral resolution of $R=7000$. This filter covers wavelengths from 648 nm to 685 nm. These observations were centered on NGC 1275 with RA : 03:19:48.16 and DEC : +41:30:42.1. Five emission lines are covered with the SN3 filter, namely : \sii$\lambda6716$, \sii$\lambda6731$, \nii$\lambda6584$, H$\alpha$(6563\AA), and \nii$\lambda6548$. The oxygen emission lines [O I]$\lambda6300$ and [O I]$\lambda6363$ cannot be observed in such a configuration since they fall just outside of the spectral range of the SN3 filter (648-685 nm). The data reduction of SITELLE is performed at CFHT through a dedicated pipeline: 
%named 
ORBS. 
%To summarize the reduction process, it involves correcting the electronic bias and the flat-field curvature of the interferometric images which are then aligned to solve potential guiding errors. 
We summarize the reduction process here. The electronic bias and the flat-field curvature of the interferometric images are first corrected. Images are then aligned to solve potential guiding errors.
Cosmic rays are 
%also
detected through an algorithm comparing successive images to determine if any abnormal flux increase can be observed for a given pixel. These detections are 
%then
corrected 
%through
 by an estimation of the given gaussian flux of nearby pixels. Atmospheric variations are also 
%considered 
 taken into account and corrected through a transmission function taken during the data acquisition. The reduction pipeline then produces the Fourier transform of all the interferograms contained within the data cube, which are then phase corrected. Finally, wavelength and flux calibrations are 
%then 
performed to adequately record the high-spectral resolution observations (\citealt{2019MNRAS.485.3930D}).

% INCLUDE paragraph about the data reduction of SITELLE at CFHT or description of SITELLE in itself for a better understanding.

\subsection{Background Subtraction}

\begin{figure}
    \centering
    \includegraphics[width=80mm,scale=0.5]{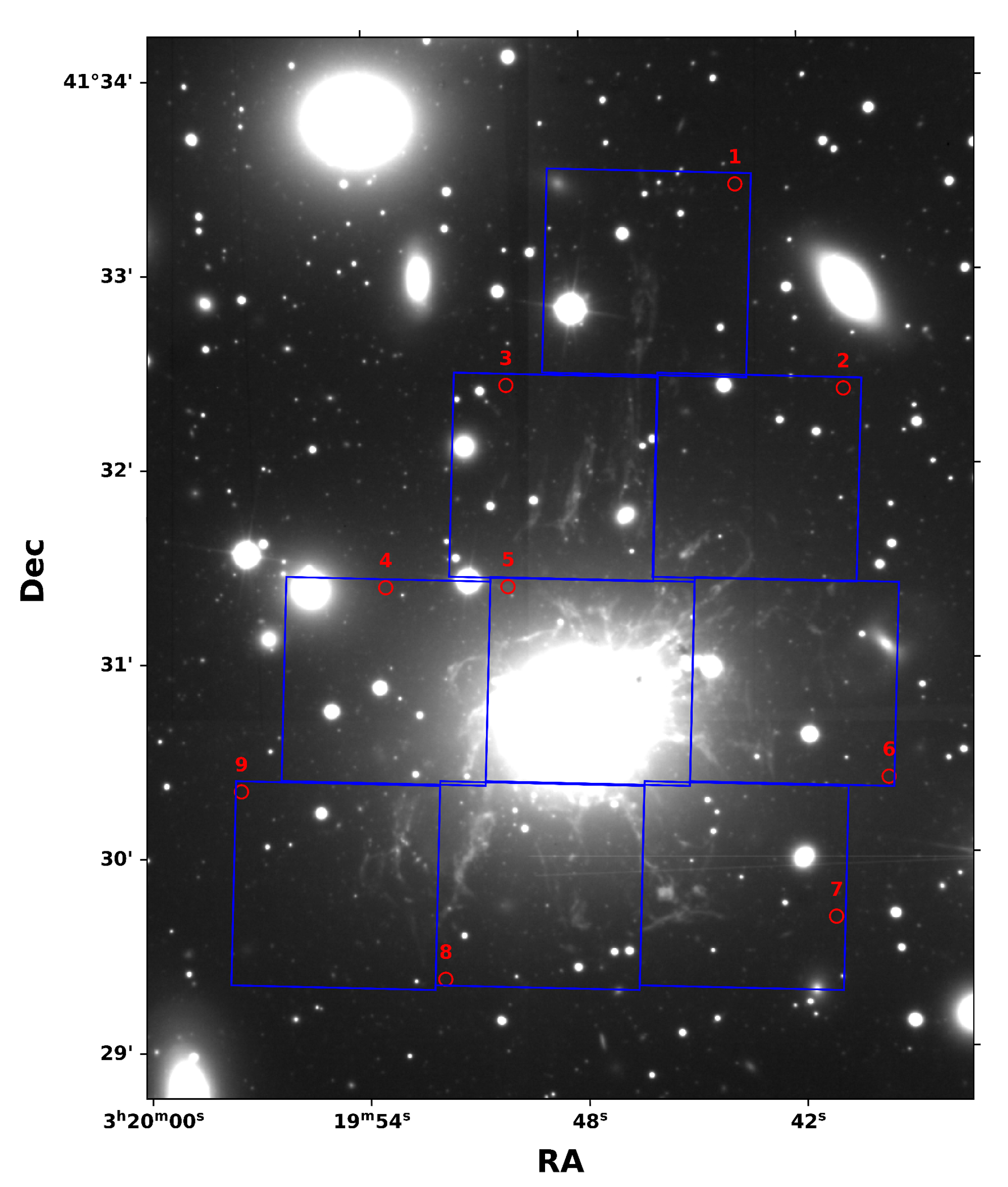}
    \caption{Mosaic map used during the analysis to tackle the background variability issue. The blue squares represent the regions whose spectra are fitted by \texttt{LUCI}, while the red circles denote the regions whose emission was used for background subtraction. The grayscale background of the figure is the deep image of the filaments surrounding NGC 1275 produced with \texttt{LUCI}.}
    \label{fig:mosaic}
\end{figure}

After a careful evaluation of the high-spectral resolution data obtained from SITELLE, it became clear that an important variability of the background emission across the field-of-view is present in the spectra (see Appendix A). This background emission is mainly produced by the galaxies present near the filamentary nebula as well as the diffuse emission of other structures in the field-of-view. Background subtraction is therefore needed to properly disentangle this unwanted emission from the filament emission lines, however, this procedure is not taken into account in the reduction pipeline.
%no proper methodology is implemented in the reduction pipeline. 
This step must therefore be performed during the fitting procedure. The background variability affects mainly the analysis of the \sii$\lambda6717$ and \sii$\lambda6731$ emission lines, which are much fainter than the other optical lines. To tackle this issue, we decided to divide the entire filamentary nebula into nine regions as illustrated in Figure \ref{fig:mosaic}. For each of these nine regions, we attributed a specific sky zone devoid of targeted emission lines and located within the boundaries of that region. The red circles shown in Figure \ref{fig:mosaic} illustrate these 9 sky zones. The sky zone is then used during the fitting procedure to perform the background subtraction for each dedicated region. Once all the regions have been fitted properly (see following sections), we can then create a mosaic of the flux, velocity, and dispersion maps for the complete filamentary nebula. This procedure allows us to counter the background variability by considering several background regions instead of one to increase the reliability of our results.
This approach allowed us to validate results obtained while considering this background variability.
The development of a dedicated methodology to tackle background variability in detail for SITELLE data will be part of a future paper.

% EXPLAIN why we do he background subtraction during the fitting routine and not in the reduction pipeline.

\subsection{Weighted Voronoi Tessellation - WVT}

From a preliminary analysis of the high-spectral resolution data of NGC 1275, we noticed that the signal-to-noise ratio (SNR) of the filamentary nebula was relatively low compared to what was expected from the SITELLE Exposure Time Calculator. Poor weather conditions during the data acquisition are considered to be the main cause of a lower SNR (see Appendix B for an example of the SNR map).

Thus, to tackle this issue, we decided to implement a weighted Voronoi tessellation algorithm as a feature of the \texttt{LUCI} software using the following tool \href{https://github.com/XtraAstronomy/AstronomyTools}{https://github.com/XtraAstronomy/AstronomyTools} (\citealt{rhea_machine-learning_2020}). This procedure creates bins of pixels whose SNR is defined by a threshold chosen by the user. Therefore, regions with strong emission will be contained within smaller bins, while noisier regions will be grouped into larger bins compared to regions of interest such as filaments. This also leads to the presence of bins containing a single pixel where the signal-to-noise ratio is already high.

More specifically, we started by making a SNR map of the filamentary nebula by only considering the H$\alpha$ emission line and \nii~emission line doublet. The SNR map is produced by determining the ratio of the maximum value of the emission lines by the standard deviation of the spectra in a waveband devoid of emission lines.
The pixels are then aggregated to create the bins and reach the SNR threshold established. In our case, we decided to fix the SNR level to 30 which is close to the estimated value of SNR = 35 for most of the filaments that would have been obtained in ideal observing conditions with SITELLE. From there, we established the number of bins defined by the algorithm and created numpy files of pixels coordinates for each to be later used in the spectral fitting procedure. The function \texttt{wvt\_fit\_region} of \texttt{LUCI} can be called to produce the fitting of a specific region of the image which will be run through the weighted Voronoi algorithm first.

\begin{figure}
    \centering
    \includegraphics[width=90mm,scale=0.5]{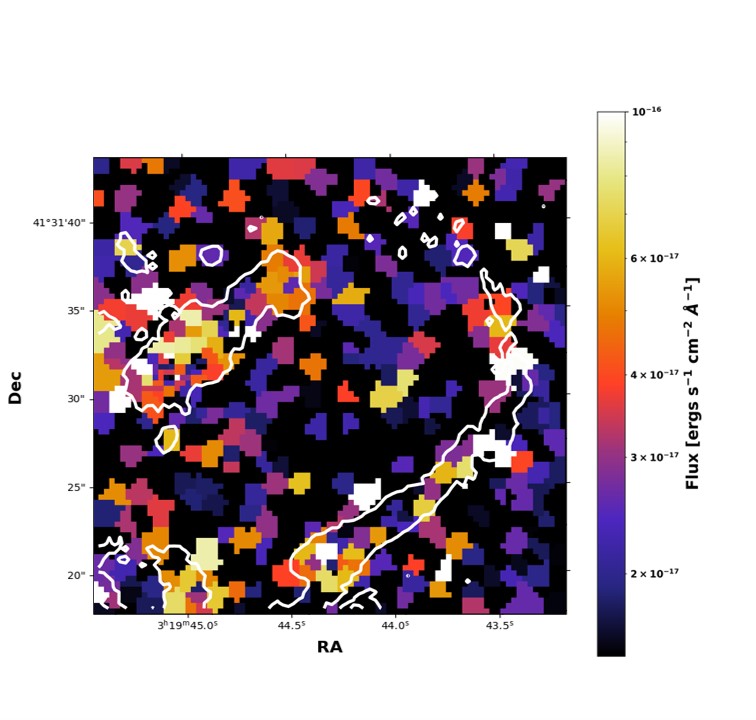}
    \caption{Example of a weighted Voronoi tesselation performed by \texttt{LUCI} to produce the resulting H$\alpha$ flux map after fitting the spectra of the horseshoe filament, whose general contour, taken from the mask displayed in Figure \ref{fig:applied_mask}, is superimposed on the figure as white contours. In this case, we had chosen a SNR value equal to 30.}
    \label{fig:wvt}
\end{figure}

One example of the WVT procedure can be seen in Figure \ref{fig:wvt} where the algorithm has been applied to the horseshoe filament as shown
%designated 
in Figure \ref{fig:fov}. It illustrates that the WVT code manages to improve the flux detection of the base of the horseshoe compared to the background emission. However, the flux
%signal
of the filament's upper region remains too low to be properly detected and accounted for in the resulting flux map. %This motivated us to then apply a flux 
We therefore applied a flux threshold of $2\times10^{-17} \text{ erg s}^{-1}\text{cm}^{-2}$ after the fitting to our final maps to 
% only consider detectable filaments. This threshold value was chosen to 
properly eliminate unwanted bins of noisy data displaying higher flux.
% in our resulting maps. 
A similar flux threshold was also applied in the analysis of low-spectral resolution observations with SITELLE by \cite{gendron-marsolais_revealing_2018}.

\begin{figure}
    \centering
    \includegraphics[width=80mm,scale=0.5]{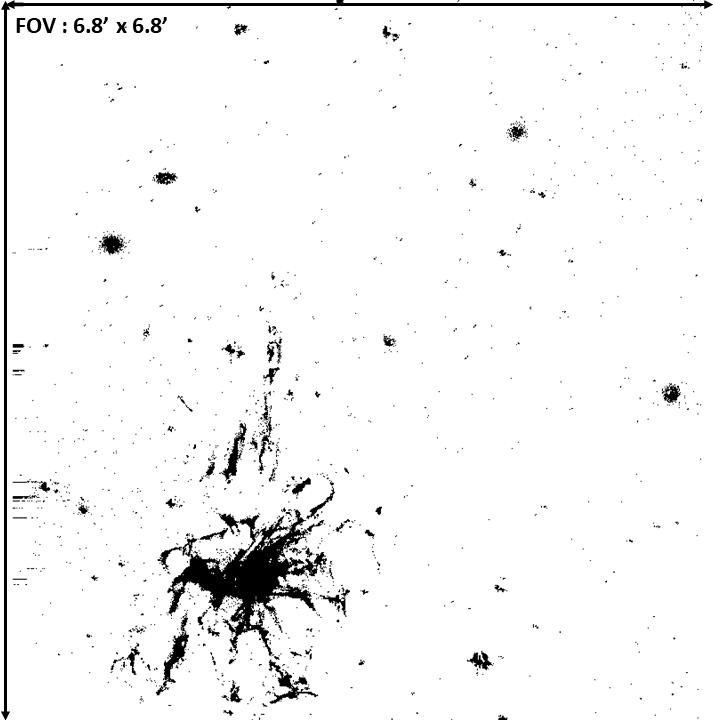}
    \caption[Masque utilisé pour les données SITELLE.]{Applied mask on the SITELLE data of NGC 1275. White pixels are masked while black pixels are unmasked. The data used to produce this mask comes from the high-spatial resolution images taken by \cite{conselice2001nature} with the WIYN telescope. Note that the field of view is smaller than the field of view of the SITELLE observations, but covers the filaments.}
    \label{fig:applied_mask}
\end{figure}

\subsection{Masking Procedure}

%To produce effective and readable maps during the fitting steps of our analysis
To help the fitting process, we decided to implement a 
%conventional 
masking procedure designed to keep as much information about the filamentary nebula as necessary. To implement this mask, we used the observational data presented by \cite{conselice2001nature} and produced by the WIYN 3.5m telescope. These deep observations showed the filamentary nebula surrounding NGC 1275 with higher signal-to-noise
%unprecedented precision and spatial resolution; 
therefore, by using these observations as a basis for our masking procedure, we can retain the necessary information for our fitting results.

The procedure used to develop this mask was implemented as follows : first, a threshold signal-to-noise value of 10 in the \cite{conselice2001nature} WIYN data was chosen in order to retain as much information regarding the filaments as possible without keeping too many background pixels in the resulting image. 
% Thus, pixels with a value below this threshold were assigned a null value while the remaining data was set to 1 to effectively create a mask with binary values that could be passed as a Boolean array (see Figure \ref{fig:applied_mask}).

Secondly, for the mask to be used in conjunction with the \texttt{LUCI} software, some groundwork was needed to perfectly match the size and position of the mask with our SITELLE data according to the WCS coordinates of the latter. Indeed, the WIYN observations did not have the same astrometry as our data, meaning that they were not properly aligned. Moreover, the number of pixels in both observations are not the same which required specific care to handle their interpolation. 

To solve this issue, we determined the number of pixels needed to cover the entirety of the SITELLE field of view by considering the pixel dimensions of both the SITELLE and WIYN instruments. Then, by knowing the position of the central pixel of the SITELLE data cube using the declination and right ascension, we were able to determine the corresponding pixel on the mask. Thus, by knowing the number of pixels needed as well as the central pixel we used the function \texttt{Cutout2D} from the \texttt{astropy.nddata} package to extend our mask to fit the SITELLE field of view. 
We then interpolated the mask onto the specific dimensions of the SITELLE observations using the function \texttt{interp2d} from the \texttt{scipy.interpolate} package, therefore creating a fitted mask without losing its nature despite having different dimensions originally. 

Finally, to properly align the mask with the SITELLE data, we used the WCS coordinates given within the SITELLE data cube to determine the position of 11 stars, which would then act as anchor points for the \texttt{estimate\_transform} and \texttt{apply\_transform} functions of the \texttt{astroalign} package. 
%Indeed, 
By determining the position of these stars within the newly produced mask, 
%we can thus match their positions 
we matched their positions to the ones of the SITELLE data and allow the overlapping of both images.

% Finally, to properly align the mask with the SITELLE data, we determined the position of 11 stars in both fields of views, which would then act as anchor points for the \texttt{estimate\_transform} and \texttt{apply\_transform} functions of the \texttt{astroalign} package.

%REFORMULATE the masking procedure to better detail the masking alignements.

\subsection{Emission Line Fitting Procedure}

After obtaining our 
%conventional
mask to the SITELLE data cube, we were able to proceed to the emission lines fitting part of the analysis. The recent software \texttt{LUCI} was used to perform the fit (\citealt{ rhea_novel_2020, rhea_machine-learning_2020, rhea_luci_2021, rhea_machine-learning_2021}) according to our mosaic structure detailed previously. Since \texttt{LUCI} is a novel analysis software, a summary of 
\texttt{LUCI} capabilities is described
%its operations and capabilities will be presented 
here but we invite the user to see \href{https://crhea93.github.io/LUCI/index.html}{https://crhea93.github.io/LUCI/index.html} for more details (\citealt{rhea_luci_2021}).

First, the spectra found in the data cube are normalized according to the highest amplitude and a shift in wavelength is applied to properly center the velocities between $-500$ 
and $500$ km/s. This constraint is necessary for the following procedure that uses machine learning since it allows better prior estimates for the fitting algorithm. This velocity range was chosen after a visual inspection of the data and considering the previous work by \cite{gendron-marsolais_revealing_2018}. These boundaries were given as training parameters for the neural network to predict the velocity shift of the emission lines. To efficiently fit a spectrum, the velocity and broadening prior information need to be precise to facilitate the minimization algorithm at the heart of the emission lines fitting and thus to accelerate the whole procedure (\citealt{zeidler_young_2019}, \citealt{bittner_gist_2019}). Therefore, a machine learning technique based on a convolutional neural network (CNN) is used to properly determine these priors. 

From there, the values obtained through the CNN are passed as the line position and broadening. The amplitude, however, is obtained from the height of the shifted emission line. With these three values, the line can be effectively fitted by \texttt{LUCI} with the \texttt{scipy.optimize.minimize} function from the \texttt{scipy} package. This function uses the Sequential Least Squares Programming (SLSQP) optimization algorithm. Since SITELLE's Instrumental Spectra Function (ISF - see \citealt{martin_optimal_2016}) is a sinc convolved with a gaussian resulting from the velocity dispersion along the line of sight, we used the sincgauss fitting model presented by \texttt{LUCI}. The results from this procedure are the amplitude, velocity, and broadening of the five emission lines present in the spectrum: \sii$\lambda6716$, \sii$\lambda6731$, \nii$\lambda6584$, H$\alpha$(6563\AA), and \nii$\lambda6548$, where we separated the fitting procedure by first considering the SNR of the H$\alpha$ and \nii\text{ } emission lines to obtain their respective parameters. Then, we specifically considered the \sii\text{ } emission lines to derive their own parameters, which will be detailed in Section 2.7.

%%%Explain which function is used for the continuuum.

Finally, to account for the location of Earth on its orbit at the time of observations, a correction is applied to the velocity map. This correction is determined with the \texttt{heliocentric\_correction} function of the \texttt{LUCI} package, and gives a velocity correction of $-27.5$ km/s.

%%% EXPLAIN which fitting function is used and how the whole procedure works.

\begin{figure}
    \centering
    \includegraphics[width=80mm,scale=0.5]{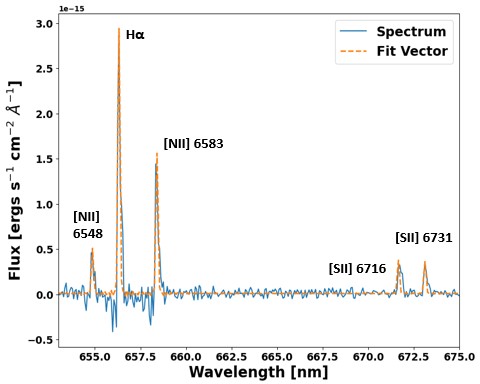}
    \caption{Example of a fitted spectrum using \texttt{LUCI} for a small circular region of $\sim 2.2$ pixels radius within the filamentary nebula surrounding NGC 1275 in the central western filaments (coordinates RA : 3:19:49.72, DEC : +41:30:51).}
    \label{fig:fitted_spectra}
\end{figure}

\begin{figure*}
\begin{subfigure}[b]{\linewidth}
  \centering
  \includegraphics[width=0.5\linewidth]{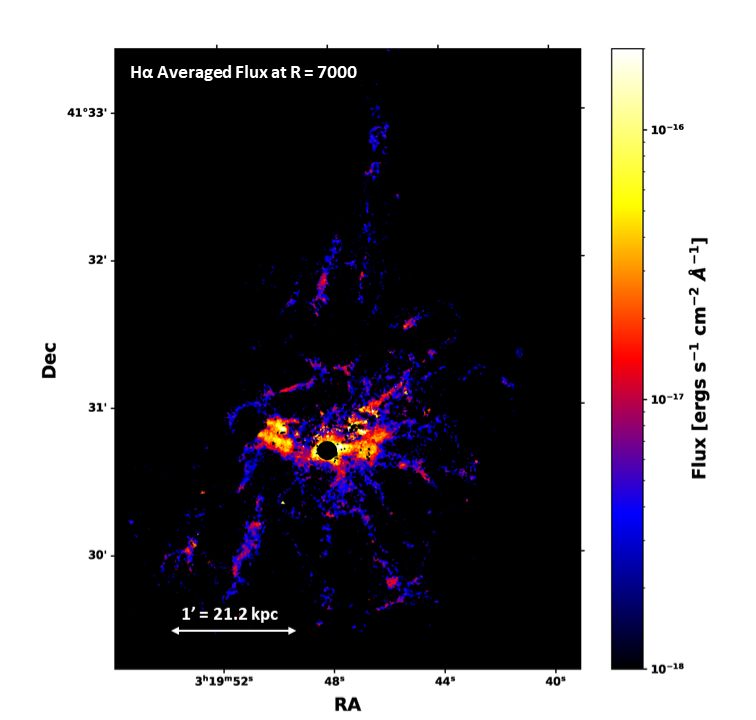}
\end{subfigure}
%\begin{minipage}[b]{0.5\linewidth}
 % \centering
  %\includegraphics[width=1.0\linewidth]{figures/Halpha_flux_map_corrected.png}
%\end{minipage}
\vskip\floatsep
\begin{subfigure}[b]{0.5\linewidth}
  \centering
  \includegraphics[width=1.0\linewidth]{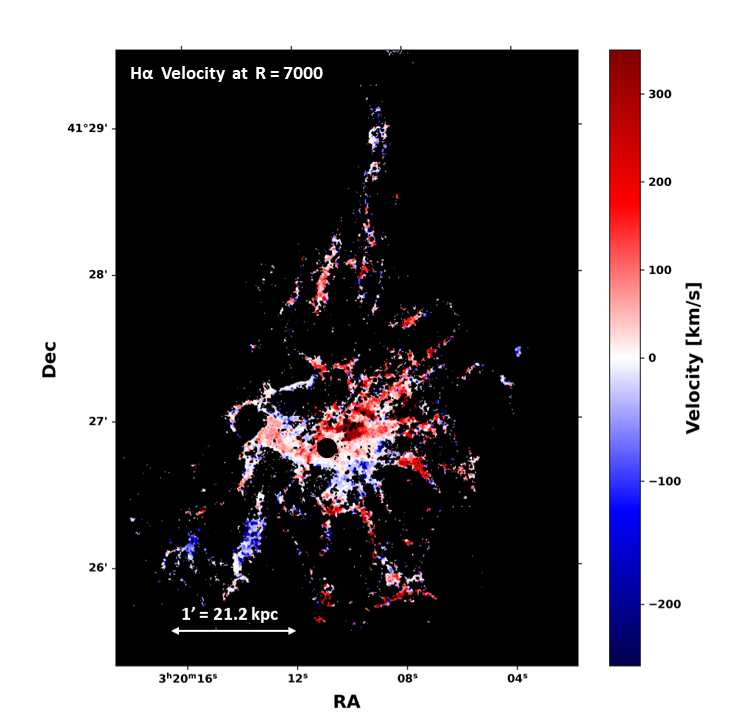}
\end{subfigure}
% \hfill
\begin{subfigure}[b]{0.5\linewidth}
  \centering
  \includegraphics[width=1.0\linewidth]{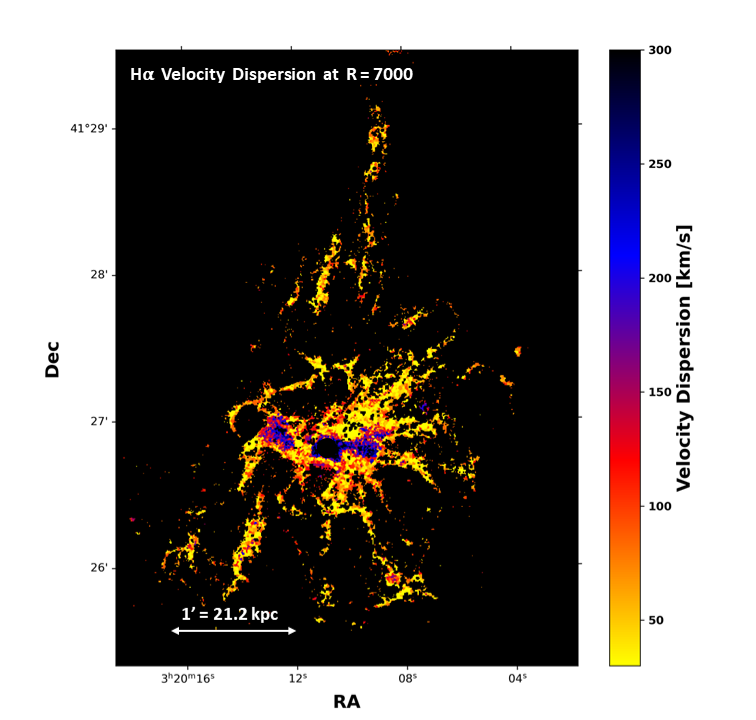}
\end{subfigure}
% \hfill
% \begin{minipage}[b]{0.5\linewidth}
%   \centering
%   \includegraphics[width=1.0\linewidth]{figures/NEW_Halpha_velocity_dispersion_map_WVT_binning.png}
% \end{minipage}
\caption{Top : H$\alpha$ averaged flux map of the filaments in NGC 1275. 
%Top right :  H$\alpha$ flux map of the filaments in NGC 1275. 
Bottom left : H$\alpha$ velocity map of the filaments in NGC 1275. Bottom right : H$\alpha$ velocity dispersion map of the filaments in NGC 1275.}
\label{fig:maps}
\end{figure*}

Through this analysis, we adopted a redshift of $z=0.017284$ for NGC 1275, which \texttt{LUCI} uses during the fitting procedure so that all calculations are done in the rest frame of the object to properly determine the velocity, velocity dispersion and flux parameters of the emission lines. An example of a fitted spectrum using \texttt{LUCI} is displayed in Fig. \ref{fig:fitted_spectra}. The resulting H$\alpha$ flux, velocity and velocity dispersion maps are shown in Fig. \ref{fig:maps}.

Finally, through these steps, we explored the fitting of averaged spectra over the bins produced by the WVT procedure to better understand the complex nature of the ionised gas in these filaments and determine their most prominent features as if no binning was applied.

%%% Include an example of a fitted spectra possibly in the Annex ???

\subsection{Multiple Emission Components}

%The high-spectral resolution observations of the filamentary nebula revealed many intricacies which required specific care during our analysis. One of these difficulties resides in the multiple velocity components that a single bin can potentially contain (see Appendix C). 
Due to the high-spectral resolution of our observations, we have to take into account the possibility of resolving multiple velocity components that a single bin can potentially contain due to the potential overlapping of filaments (see Appendix C). Indeed, since the three-dimensional structure of the filaments is currently unknown, it is impossible to decipher the possible overlap of several filaments. 

% The forefront galaxy also acts as a source of contamination in our spectra. 

For the sake of clarity in our resulting maps, we decided to mask the main sources of multiple emission components. Firstly, the high-velocity system, associated with a forefront galaxy, has a velocity of $\sim 8200$ km/s. It can thus be identified through its systemic velocity and ignored during the fitting procedure by only considering the emission lines associated with the filamentary nebula.

\begin{figure}
    \centering
    \includegraphics[width=80mm,scale=0.5]{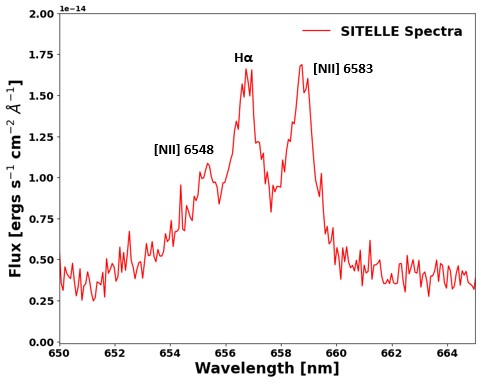}
    \caption{Example spectrum extracted from the central region of the filamentary nebula surrounding NGC 1275 centered on the AGN. Narrow emission lines overlaid on a broad component are clearly seen. This region has a radius of $\sim 5$ pixels.}
    \label{fig:AGN_spectra}
\end{figure}

Moreover, broad, multiple component emission lines from the AGN are coupled with the filament emission in the central $\sim 2.6$ kpc, thus, the resulting fits produced by \texttt{LUCI} are poor when considering these specific regions. An example of a spectra extracted from the central region close to the AGN is displayed in Fig. \ref{fig:AGN_spectra}.

% The AGN located at the center of NGC 1275 is also a powerful source of emission with extremely broad emission lines. 

The large broadening of the AGN's emission overlapping with the filaments' narrow emission impedes \texttt{LUCI} from properly fitting the central regions of the filamentary nebula. Indeed, the fitting algorithm proves to be limited by the presence of largely broadened and blended emission lines for which the machine learning algorithm has not been trained on, therefore preventing it from performing accurate fitting. Therefore, to facilitate the analysis, we decided to mask the central region of emission coming from the AGN displaying broad emission lines. To do so, we studied the emission line profiles with \texttt{LUCI} and determined that the AGN spectral presence is confined within a radius of $\sim 2.6$ kpc from the center of the galaxy. There was also evidence of a small number ($\sim 10$) of localized clumps detected over 1 to 5 pixels within the central regions ($r<10$ kpc) that showed evidence of multiple components in the emission lines. Similar clumps had already been detected in a previous study by \cite{hatch_origin_2006}, where the authors mentioned the presence of several regions displaying double-peaked emission lines.

%After a careful examination of the SITELLE spectra based on the previous analysis of \cite{hatch_origin_2006}, we detected several regions presenting multiple velocity components all found in the central part of the filamentary nebula within $r<10$ kpc. 
Some of these regions appear as small knots or plumes displaying an extremely low separation in velocity components, leading to slightly double-peaked emission lines. However, we can also detect slightly larger regions displaying several components in velocity and located near the eastern part of the filamentary nebula below the high-velocity system. It is not yet clear if these regions with multiple velocity components are due to the superposition of filaments with Doppler-shifted velocities in the line of sight. Spectra of specific regions 
%of interest 
displaying several emission lines components are presented in Appendix C.

%%% EXPLAIN which type of BCG NGC 1275 is and why the broad emission lines are difficult to fit. They presence prove difficult for LUCI to fit and produce meaningful results.

%%% Show an example of the fitted central region to display the broad component.

%%% Show velocity maps between -300 and 300 km/s (different than ML maps) ???

Here we focus only on the results from the fitting of a single component to all bins. A dedicated multiple-component analysis of the few localized regions with overlapping filaments, as well as of the core, will be performed in the future.

%Since these are rare and localized only in the core of the galaxy, we decided to only fit a single component to all bins in order to extract their spectral information in this paper. An extended analysis will be performed in the future with a dedicated procedure to adequately fit these closely related emission components, only displaying small shifts in velocity, as well as developing a joint model to properly disentangle the broad AGN emission in the central region of the filamentary nebula.

\subsection{[SII] Emission Lines Fitting}

Regarding the detection and fitting of the \sii$\lambda6716$ and \sii$\lambda6731$ emission lines, one of the main difficulties in their analysis resides in the presence of strong sky lines at the same wavelengths (see Appendix A).
%in the fact that sky lines falls near them 
%(see Appendix A)
However, after applying our methodology to tackle the background variability, we were able to detect %and resolve
these emission lines. 
%To help the fitting procedure, we made new WVT bins based on the SII SNR doublet, which are fainter than the H alpha and NII lines.
To help the fitting procedure, we made new WVT bins based on the \sii$\lambda6716$ and \sii$\lambda6731$ doublet SNR, which are fainter than the \nii$\lambda6583$, H$\alpha$(6563\AA), and \nii$\lambda6548$ lines.

%Once this was done, to improve the fitting procedure made with \texttt{LUCI} even more, some additional changes were implemented.

\begin{figure*}
\begin{subfigure}{0.5\textwidth}
    \centering
    % include first image
    \includegraphics[width=1.0\linewidth]{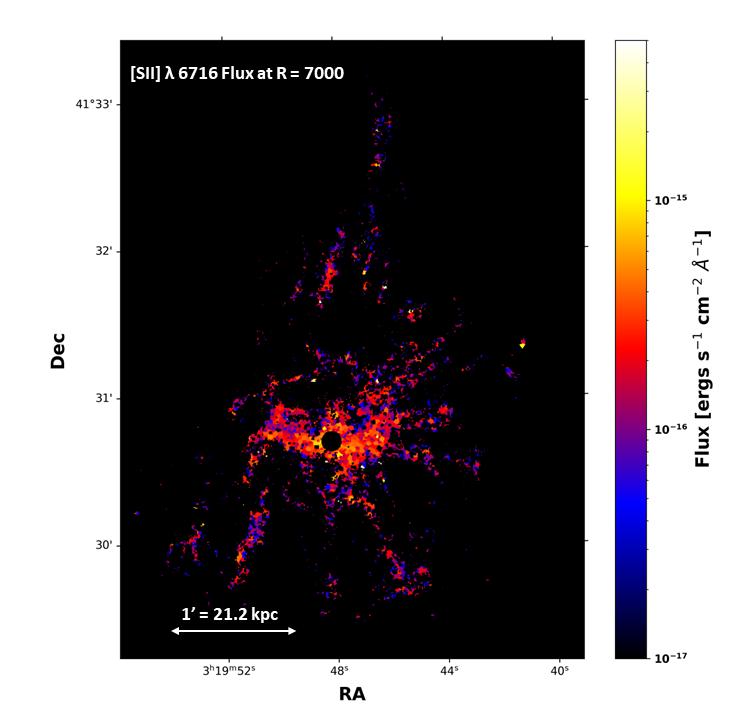}  
    %\label{fig:sub-first}
\end{subfigure}%
\begin{subfigure}{0.5\textwidth}
    \centering
    % include second image
    \includegraphics[width=1.0\linewidth]{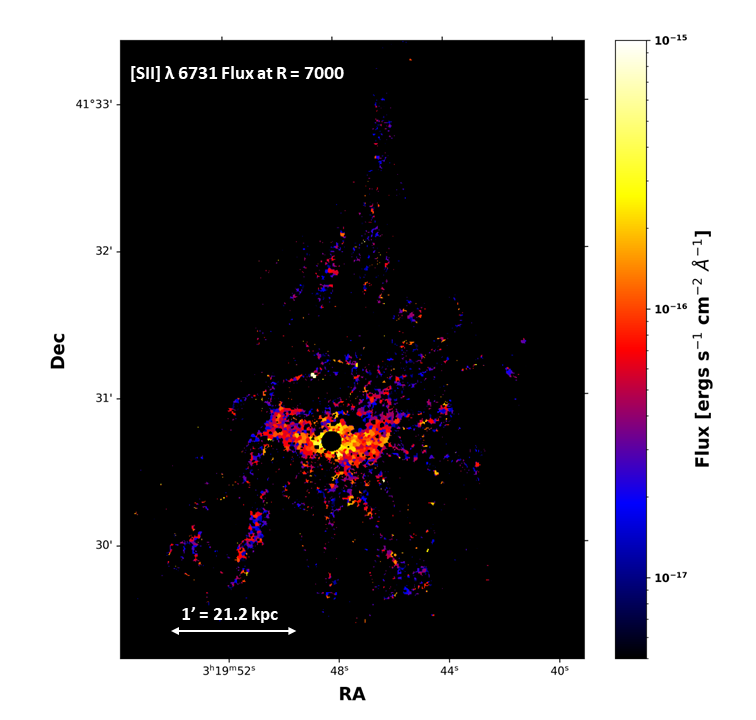}  
    %\label{fig:sub-second}
\end{subfigure}
\newline
\begin{subfigure}{0.5\textwidth}
    \centering
    % include third image
    \includegraphics[width=1.0\linewidth]{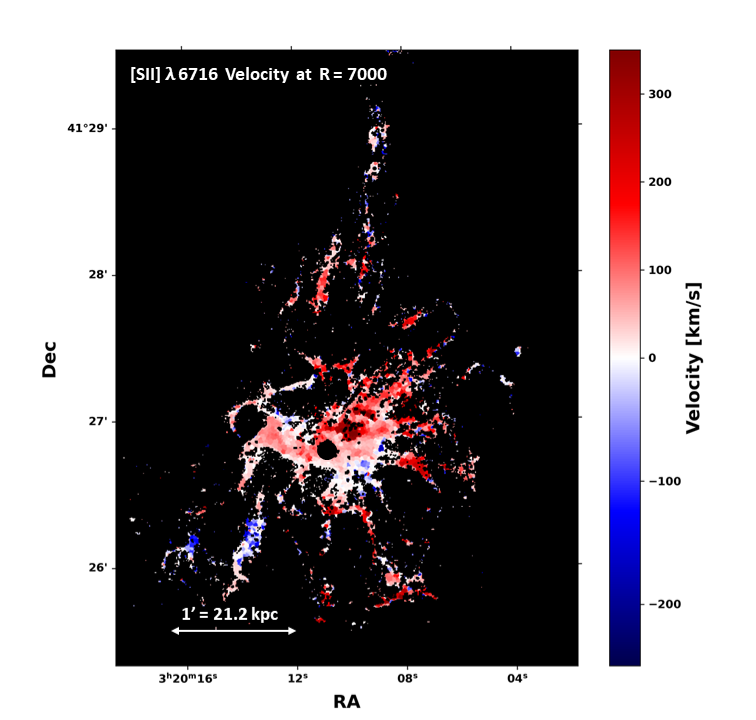}  
    %\label{fig:sub-third}
\end{subfigure}%
\begin{subfigure}{0.5\textwidth}
    \centering
    % include fourth image
    \includegraphics[width=1.0\linewidth]{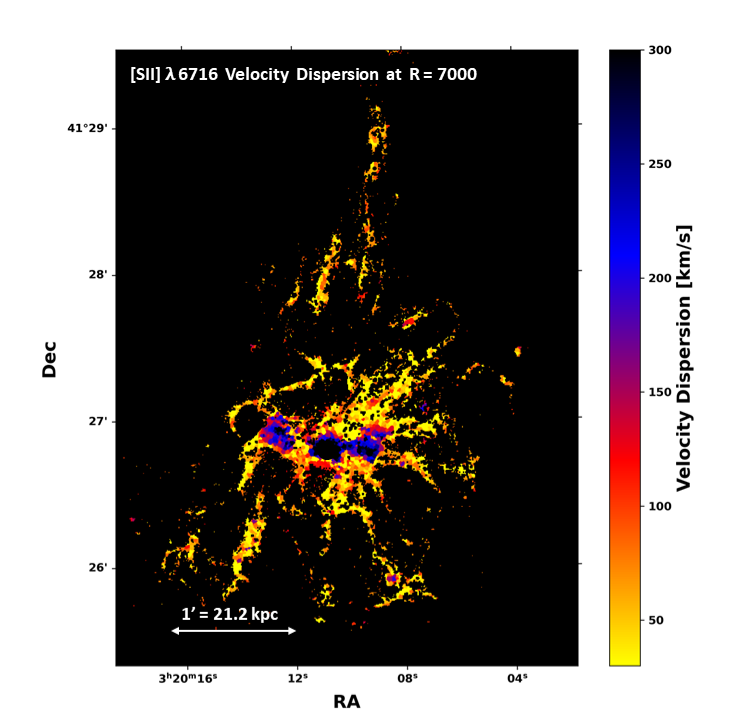}  
    %\label{fig:sub-fourth}
\end{subfigure}
\caption{Upper left: \sii$\lambda6716$ flux map of the filaments in NGC 1275. Upper right: \sii$\lambda6731$ flux map. Bottom left: \sii$\lambda6716$ velocity map. Bottom right: \sii$\lambda6716$ velocity dispersion map. The velocity and velocity dispersion of \sii$\lambda6731$ are the same as the ones for the \sii$\lambda6716$ line since these parameters were tied when fitting the doublet.}
\label{fig:SII_maps}
\end{figure*}

%Indeed, since the \sii$\lambda6716$ and \sii$\lambda6731$ emission lines are fainter than the \nii$\lambda6583$, H$\alpha$(6563\AA), and \nii$\lambda6548$ lines, we decided to build the WVT bins based on the SNR of the \sii\text{ }doublet instead. 
The new larger bins thus created improved the detection of these fainter emission lines.
%This led to larger bins and improved the general detection of these emission lines. 
%This has allowed us to produce the first analysis of the ionised sulfur emission in the entirety of the filamentary nebula, which will be detailed in section 3.8.
%Afterwards, in order to produce the ratio map of \sii$\lambda6716$ by H$\alpha$(6563\AA) (see bottom left plot of figure \ref{fig:ratio_maps}), we specifically applied the fitting procedure for the H$\alpha$ emission line by considering the bins obtained through the WVT algorithm which uses the SNR of the \sii\text{ } doublet. Thus we can produce a flux ratio map for both the emission lines with the exact same number of bins. 
The resulting \sii$\lambda6716$ and \sii$\lambda6731$ flux maps, velocity and velocity dispersion maps are shown in Fig. \ref{fig:SII_maps}. The bottom left and right plots clearly show similar velocity and velocity dispersion maps as H$\alpha$ seen in Fig. \ref{fig:maps}. A deeper analysis of the \sii\text{ }emission lines ratio, which are often used as an estimate of the gas density compared to the density of the hot X-ray gas, will be presented
%performed
in a future article (Vigneron et al., in preparation).

\section{Results and discussion} \label{sec:displaymath}

\subsection{Comparison with Previous SITELLE Observations}

\begin{figure*}
    \centering
    \includegraphics[width=180mm,scale=0.5]{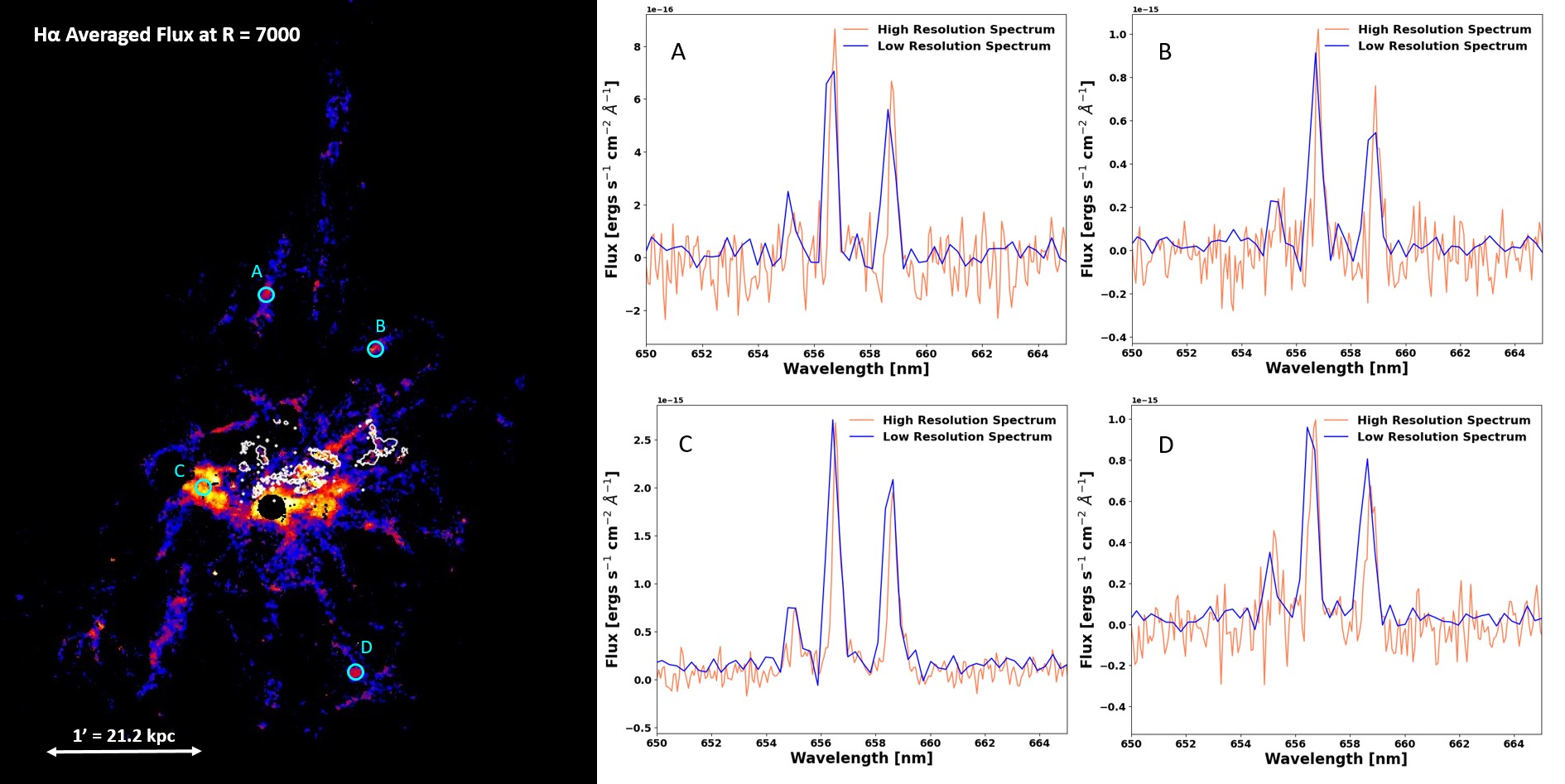}
    \caption{Comparison of the spectra extracted from four regions denoted by the letters A, B, C, and D for both low and high-spectral resolution. On the left side, the high-spectral resolution H$\alpha$ averaged flux map is used to display the region of emission while the right side showcases the differences between low and high-resolution spectra.The white contours overlayed on the left-hand flux map display the HVS structure situated in the forefront of the filamentary nebula of NGC 1275.}
    \label{fig:spectra_comparison}
\end{figure*}

Previous SITELLE R=1800 observations of the filamentary nebula surrounding NGC 1275 were already analyzed by \cite{gendron-marsolais_revealing_2018} and revealed several key features of the velocity structure.

Our new high-spectral resolution observations, however, presented new challenges during the analysis step. Indeed, the spectra produced by SITELLE uses a Fourier transformation based on a sinc function, which shows specific lobes around the central peak. 
%For low-spectral resolution data, these lobes are present but tend to be unclear spectral features. On the other hand, our new high-spectral resolution observations are better suited to display these spectral lobes. Indeed, Figure \ref{fig:spectra_comparison} shows the comparison of spectra extracted from four regions of the filamentary nebula at low and high-spectral resolution. 
Since SITELLE instrumental function is a sinc function convoluted with a gaussian resulting from the velocity dispersion along the line of sight, a sincgauss model was specifically chosen during the spectral fitting procedure with \texttt{LUCI} to properly take into account the spectral information contained within the lobes
%, these lobes contain spectral information, thus, they were taken into account during the spectral fitting procedure with \texttt{LUCI} by specifically choosing a sincgauss model.

Moreover, the spectral resolution associated with the SITELLE observations induce a minimum value of velocity dispersion that can be resolved clearly (see Figure 3 in \citealt{2019MNRAS.489.5530R}). Thus, when considering a small spectral resolution of R = 1800 with the SITELLE instrument, as was the case for the observations studied in \cite{gendron-marsolais_revealing_2018}, the lower velocity dispersion bound that could be obtained was of $\sim 80$ km/s. However, with a high-spectral resolution of R = 7000, the lower velocity dispersion value that can be determined is of $\sim 15$ km/s, allowing us to properly resolve the emission lines (see Figure 3 of  \citealt{2019MNRAS.489.5530R}). This result is visible in Figure \ref{fig:spectra_comparison} when comparing the spectra of the same region at both low and high-spectral resolution. We can indeed observe that the width of the emission lines is smaller than what was previously found by \cite{gendron-marsolais_revealing_2018} in region A, B and D. Region C is contained within the central region and given its intrinsic large velocity dispersion, the low-spectral resolution observations of SITELLE were capable of resolving this structure. 

For illustrative purposes, we also overplotted the contours of the HVS on the left-hand map of Figure \ref{fig:spectra_comparison} These contours were created by forcing LUCI to specifically fit the emission line from the HVS thus allowing us to isolate it on a dedicated small region of the SITELLE data.

\subsection{Emission Line Intensity Ratios and Flux Maps}

%Through our analysis of the high-spectral resolution data with \texttt{LUCI}, we have obtained several resulting maps that will be presented throughout this section. This subsection will specifically discuss the flux and ratio maps as well as the information they convey regarding the ionization properties involved in the filamentary nebula.

First, the H$\alpha$ averaged flux map visible in the upper plot of Figure \ref{fig:maps} shows a disk-shaped structure in the central region with a size of $\sim 22$ kpc by $5$ kpc displaying a flux an order of magnitude higher than the rest of the filaments. The flux map from \cite{gendron-marsolais_revealing_2018} hinted to this structure, but the new data reveal it in much greater detail. The high averaged flux of this disk-shaped structure is also associated with a large velocity dispersion, which will be discussed in section 3.4.
Beyond the disk-shaped structure, localised parts of specific filaments display slightly higher H$\alpha$ fluxes, but most of them have a homogeneous lower flux. 
%\textbf{When considering the upper right non-averaged flux map in Figure \ref{fig:maps}, we can see that most of the emission is localized closer to the central region but the outer filaments appear more visible and detectable as was described with the WVT method.}

\begin{figure*}
\begin{subfigure}{0.5\textwidth}
    \centering
    % include first image
    \includegraphics[width=1.0\linewidth]{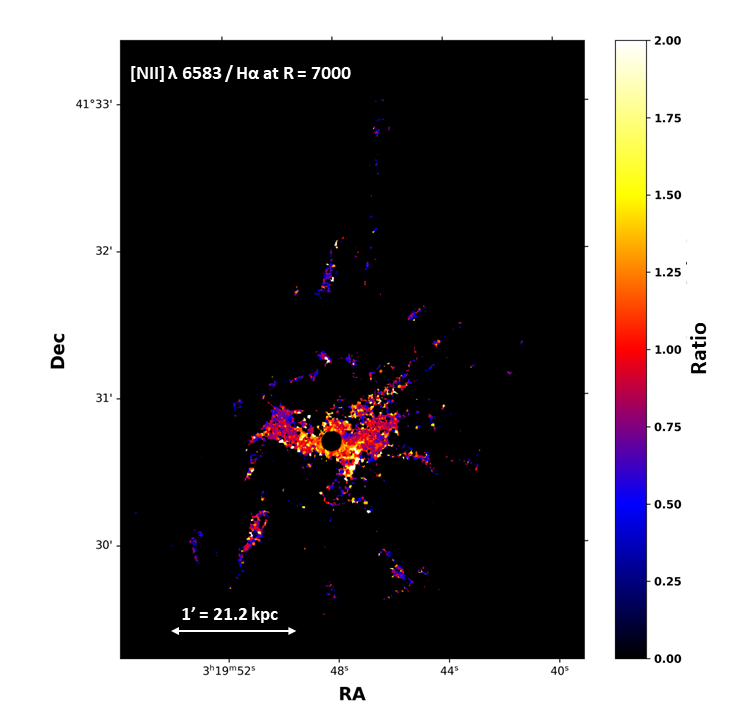}  
    %\label{fig:sub-first}
\end{subfigure}%
\begin{subfigure}{0.5\textwidth}
    \centering
    % include second image
    \includegraphics[width=1.0\linewidth]{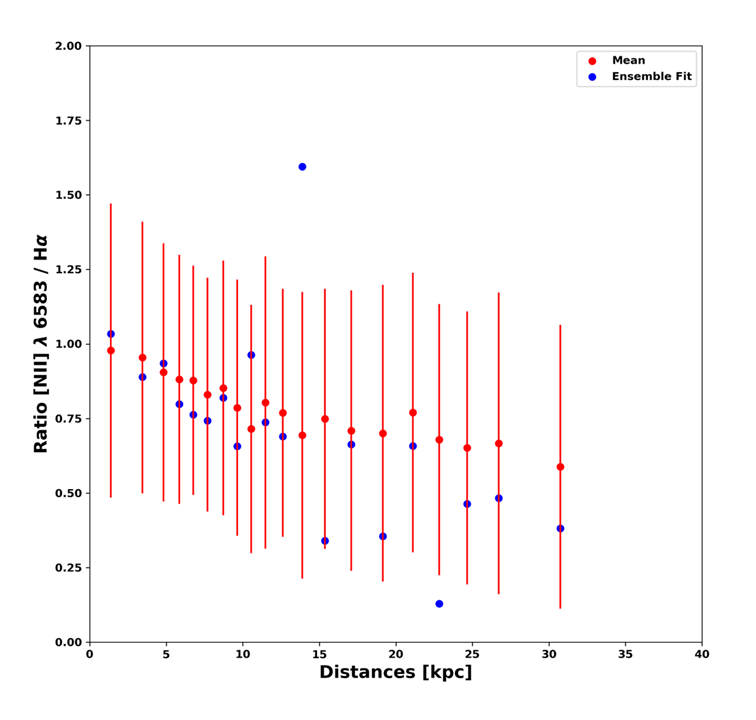}  
    %\label{fig:sub-second}
\end{subfigure}
\newline
\begin{subfigure}{0.5\textwidth}
    \centering
    % include third image
    \includegraphics[width=1.0\linewidth]{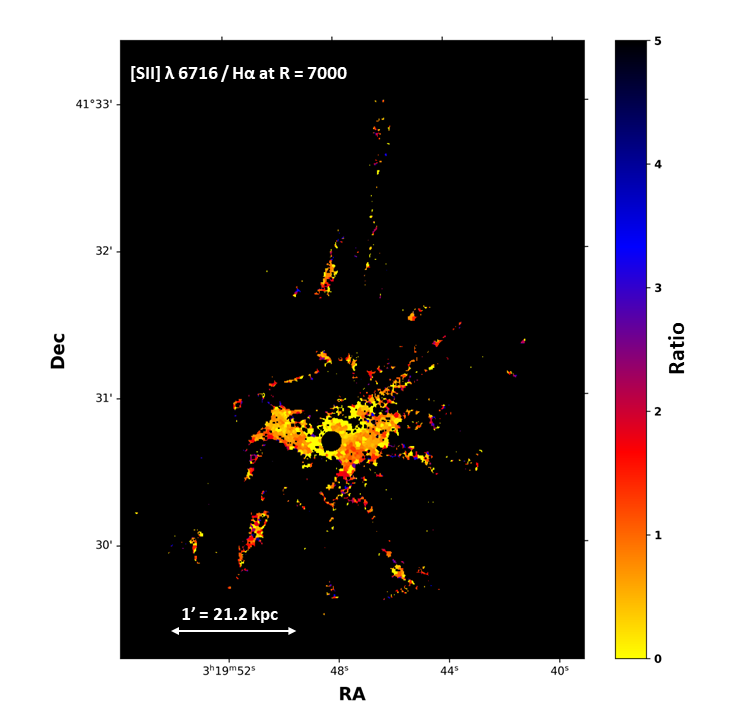}
    %\label{fig:sub-third}
\end{subfigure}%
\begin{subfigure}{0.5\textwidth}
    \centering
    % include fourth image
    \includegraphics[width=1.0\linewidth]{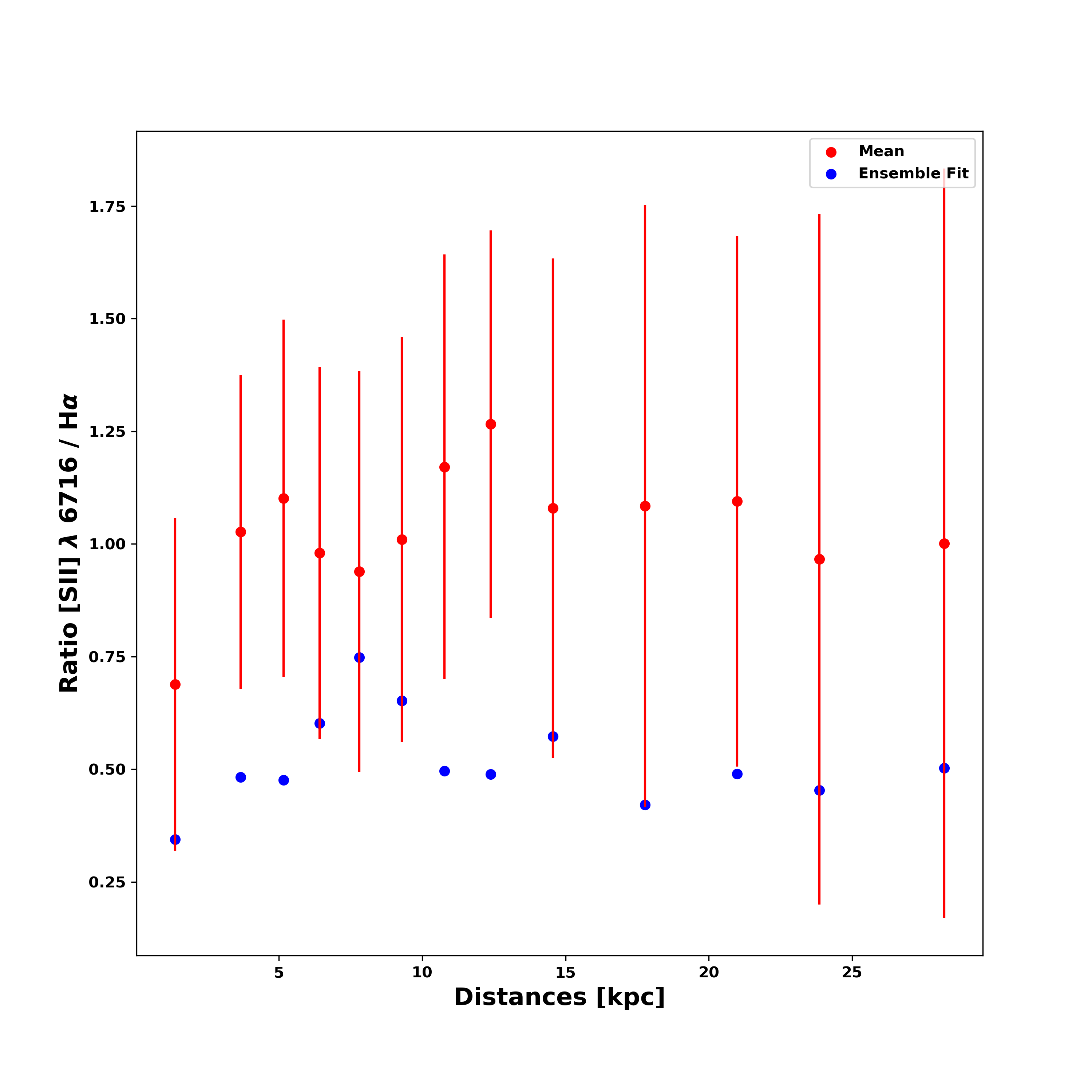}  
    %\label{fig:sub-fourth}
\end{subfigure}
\caption{Upper left: \nii$\lambda6583$ by H$\alpha$ ratio map. Upper right: Mean (in red) and ensemble fits (in blue) across annuli containing 1500 pixels for the \nii$\lambda6583$ by H$\alpha$ ratio. The error bars for the ensemble fit have been plotted but are too small to see compared to the data scale. Bottom left: \sii$\lambda6716$ by H$\alpha$ ratio map. Bottom right: Mean (in red) and ensemble fits (in blue) across annuli containing 1500 pixels for the \sii$\lambda6716$ by H$\alpha$ ratio. The error bars for the ensemble fit have been plotted but are too small to see compared to the data scale.}
\label{fig:ratio_maps}
\end{figure*}

Emission line ratios can be of great interest when studying the ionization properties of an emitting gas (see \citealt{kewley2019understanding}). In this regard, the \nii$\lambda6583$/H$\alpha$ ratio is of importance since it informs us of the relative implication of soft versus hard sources of ionization. In the case of a low ratio (meaning H$\alpha$ flux $\gg$ \nii$\lambda6583$ flux), photons from young stars can be the ionizing source, while for a higher ratio (meaning H$\alpha$ flux $\leq$  \nii$\lambda6583$ flux), a harder ionizing spectrum (from an AGN, for example) or energetic ICM particles are needed to provide additional heat
%a higher level of ionisation is needed, thus, AGN activity or energetic ICM particlesare more likely to ionise the gas 
(see \citealt{article}). Similarly, the ratio of \sii$\lambda6716$ by H$\alpha$(6563\AA) can be used as a confirmation value for various ionization mechanisms. Here, to produce the ratio map of \sii$\lambda6716$ by H$\alpha$(6563\AA), we specifically applied the fitting procedure for the H$\alpha$ emission line by considering the bins obtained through the WVT algorithm which uses an SNR of 35 for the \sii\text{ } doublet (between wavelengths of 671,1 nm to 675,7 nm) thus creating bins of the exact same size and position as the \sii\text{ } doublet flux maps but fitting the H$\alpha$ emission line at $\lambda = 656$ nm. Afterwards, we can obtain the ratio map by dividing both these flux maps therefore producing a flux ratio map for both the emission lines with the exact same number of bins (see bottom-left plot of Figure \ref{fig:ratio_maps}).

% which uses the SNR of the \sii\text{ } doublet. Thus we produced a flux ratio map for both the emission lines with the exact same number of bins.}

%Therefore, in the case of the filamentary nebula surrounding NGC 1275, we can see in the upper left plot of Figure \ref{fig:ratio_maps} 
Figure \ref{fig:ratio_maps} upper-left map shows that the central region of the structure displays a ratio of \nii$\lambda6583$/H$\alpha$ close to or slightly superior to 1.0, while the extended filaments show a variety of smaller ratio values between 0.5 to 1.0. This trend between central and extended filaments can also be seen in the upper right plot of Figure \ref{fig:ratio_maps}, where we produced the mean (i.e. taking the mean of the ratio values accross the annuli) and ensemble fit results (i.e. extracting one combined spectra for the entirety of the pixels inside the annulus and fitting it to obtain an ensemble ratio value) for the \nii$\lambda6583$/H$\alpha$ ratio across annuli containing 1500 pixels each, reproducing \cite{gendron-marsolais_revealing_2018}. 
However, in the optics of emulating this methodology for high spectral resolution SITELLE data, we noticed that ensemble spectra at high spectral resolution are severely affected by the spatial distance between annuli pixels leading to blended emission lines and to large offsets between the mean and ensemble fits. We have decided to leave those results, but we are analyzing only the mean fit results.
%preventing us from reaching similar conclusions as those obtained from an analysis at lower spectral resolution (see \citealt{gendron-marsolais_revealing_2018} for effective ensemble fit profiles at a lower spectral resolution). 
%However, they do serve as an interesting methodology comparison between similar SITELLE observations at different spectral resolution.}

The center of each annuli is determined based on the WCS coordinates of the central galaxy NGC 1275 given as RA : 03:19:48.16 and DEC : +41:30:42.1 through the literature (see \citealt{2020yCat.1350....0G}) and observed in our SITELLE data. The width of the annuli is determined according to the fixed number of 1500 detected pixels they contain. Therefore, they vary between widths of 7 to 10 pixels and up to 57 pixels for the outer filaments.
Error bars are also plotted for both methods and obtained 
%either 
through statistical calculations for the mean, or through \texttt{LUCI}'s fitting procedure for the ensemble fit. The error bars for the ensemble fit have also been plotted but are too small to see compared to the data scale. This profile shows a decrease in the mean values from the center (0 - 10 kpc) to the outer filaments (10 - 30 kpc) from $1.0 \pm 0.5$ to $0.6 \pm 0.5$, thus suggesting a gradual change or variation of the ionisation mechanism. However, the ratio values appear more chaotic than structured in the extended filaments which is reflected by a fairly large standard deviation in the error bars of the mean profile (see upper right plot of Figure \ref{fig:ratio_maps}).

%This result 
Such radial trend is similar to what was previously found by \cite{gendron-marsolais_revealing_2018} with the low-spectral resolution SITELLE data (R=1800), but also through slit spectroscopy (\citealt{hatch_origin_2006}). The presence of star formation has already been detected for specific southern and north-western regions of the filamentary nebula (see \citealt{canning2010star, canning_filamentary_2014}). However, photoionization models predict a ratio that is highly dependent on metallicity, with an upper limit of around 0.5 (\citealt{kewley_host_2006}), thus, such a process could not explain the higher ratios found throughout the southern filaments. In addition, we can also conclude that the central region, near the AGN, is more likely to be ionised by its activity as well as energetic particles since it displays higher ratios, while the outer filaments could predominantly be ionized by energetic particles thus showing intermediate \nii$\lambda6548$ / H$\alpha$ emission line ratios.

Another possibility that is often presented as a mechanism to explain the energetics and ionisation of the filaments is through collisional excitation by energetic particles coming from the hot ICM (see \citealt{ferland_collisional_2009}). In such circumstances, the \nii$\lambda6548$ / H$\alpha$ emission line ratio would be close to $\sim 0.3$. However, if we consider the ratio of \sii/H$\alpha$, displayed in the bottom left plot of Figure \ref{fig:ratio_maps}, we can see that the outer filaments show higher ratios above $\sim 1$, nevertheless, various behavior are observed for different filaments. In the case of ionization by cosmic rays, the ratio of \sii/H$\alpha$ is predicted to be at $1.4$ according to the cosmic ray heating model (see Table 5 of \citealt{ferland_collisional_2009}), which could be an argument in favor of the collisional ionisation model for the outer filaments. However, the northern filaments as well as parts of some other filaments show even higher \sii/H$\alpha$ ratio values above $\sim 2.0$. Nevertheless, in the bottom right plot of Figure \ref{fig:ratio_maps}, we produced the mean and ensemble fit results of the \sii$\lambda6716$/H$\alpha$ ratio across annuli containing 1500 pixels each, we can see that most mean values are constrained between $\sim 0.5$ and $1.5$ which could argue in favor of an ionization mechanism through cosmic rays, though it is not possible to offer conclusive results with the emission lines at hand. 

Using subsequent SITELLE observations with SN1 (365 - 385 nm) and SN2 (480 - 520 nm) filters of NGC1275 (PI: G.-Marsolais), that provides other emission lines, we are currently pursuing a deeper analysis of the ionization mechanisms at play in the filamentary nebula (Rhea et al. in preparation).
%Finally, to pursue a more detailed analysis of the ionization mechanisms involved in the filamentary nebula, other line ratios are needed. By analyzing subsequent observations with the SN1 (365 - 385 nm) and SN2 (480 - 520 nm) filters of SITELLE, we would have the necessary line ratios to perform complete Baldwin-Phillips-Terlevich diagrams (BPT, \citealt{baldwin_classification_1981}). These filters would indeed allow us to obtain the H$\beta\lambda4861$, [O III]$\lambda4959$, [O III]$\lambda5007$, [O II]$\lambda3726$ and [O II]$\lambda3729$ emission lines. This work is currently underway and will be published in a future article (Rhea et al. in preparation). 
Similarly, the analysis of \sii~emission lines and their ratios will be explored in a future article (Vigneron et al. in preparation)

%%% For the annuli figures, describe how we choose the center of the annuli

\subsection{Central disk-shaped Structure}

%%% REPHRASE mentions of the radio bubbles or show their contours on the SITELLE maps.

%%% ARE the results more focused on the central structure, if so, the mention should be made clear in the introduction.

%%% MENTION NIFS observations that display an inner rotating disk structure. 

\begin{figure*}
\begin{subfigure}{0.5\textwidth}
    \centering
    \includegraphics[width=0.95\linewidth]{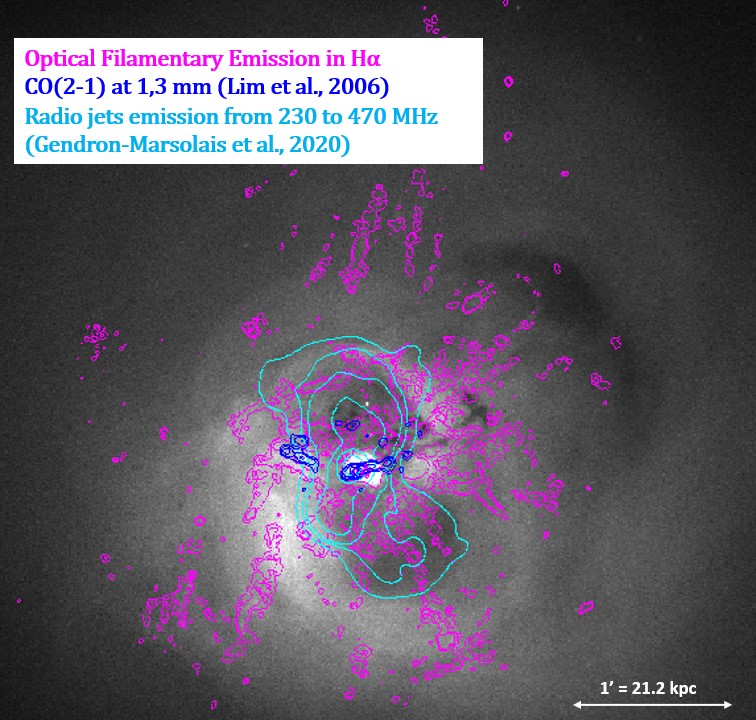}  
\end{subfigure}%
\begin{subfigure}{0.5\textwidth}
    \centering
    \includegraphics[width=0.95\linewidth]{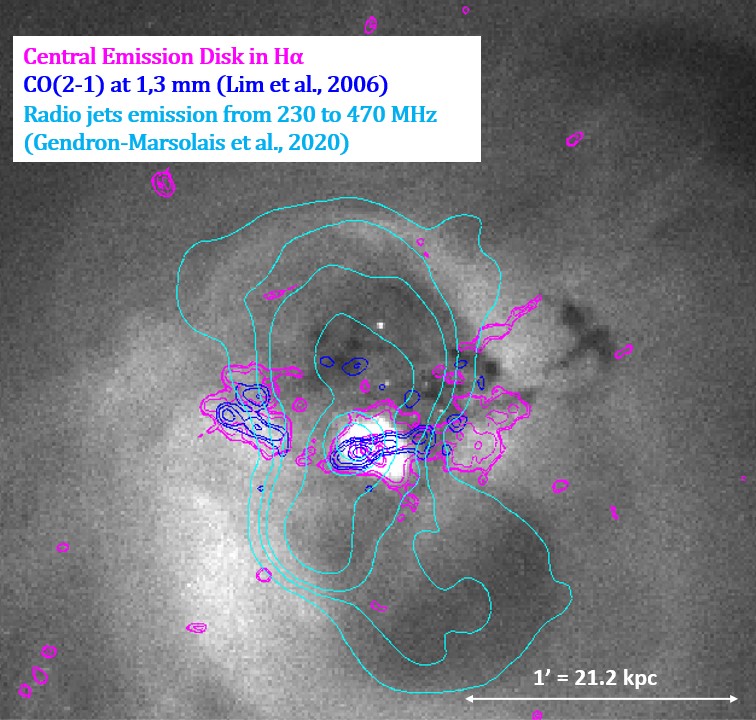}  
\end{subfigure}
\caption{Left: Chandra 0.5 - 2.0 keV observations of NGC 1275.
The magenta contours show \color{black}
%illustrate
the H$\alpha$ 
%emission 
filaments as observed with SITELLE at high-spectral resolution at a flux level stating at $1\times10^{-18} \text{ erg s}^{-1}\text{cm}^{-2}\text{\AA}^{-1}$. The cyan contours display the radio emission from 0.23 to 0.470 GHz as observed with the VLA (\citealt{gendron-marsolais_high-resolution_2020}) 
%thus 
tracing the radio jets. The blue contours represent the CO(2-1) central emission disk-shaped structure as observed with the SMA (\citealt{lim_radially_2008}). Right: Zoomed-in image of the left-side figure with a H$\alpha$ flux cut-off of $1\times10^{-17} \text{ erg s}^{-1}\text{cm}^{-2}\text{\AA}^{-1}$.}
\label{fig:spatial_correlation}
\end{figure*}

As we mentioned, the filamentary nebula seen in the optical is part of a larger multiphase structure that correlates spatially with filament-like structures seen in X-rays (\citealt{fabian_relationship_2003}, \citealt{walker_x-ray_2015})
%, Hitomi \citealt{hitomi_collaboration_atmospheric_2018}) 
and molecular gas structures 
%in radio
(\citealt{salome_cold_2008}, \citealt{2009ApJ...698.1191H})
%, \citealt{gendron-marsolais_high-resolution_2020}). 
%These emissions at different wavelengths display a clear spatial correlation between each other and 
Such multi-phase filaments are at the heart of many models explaining the formation of such structures. Indeed, it is argued that the black hole jets create expanding bubbles, rising through and carving the ICM, while inducing turbulence, sound waves and shocks (e.g. \citealt{dhawan_kinematics_1998}, \citealt{falceta-goncalves_turbulence_2010}). This effect can be seen through the deformation of optical filaments taking the shape of convection cells and their spatial correlation with the trailing radio bubbles (\citealt{fabian_relationship_2003}; see also the horseshoe filament in left plot of Fig \ref{fig:spatial_correlation}).

\begin{figure*}
\begin{subfigure}{0.5\textwidth}
    \centering
    % include first image
    \includegraphics[width=1.0\linewidth]{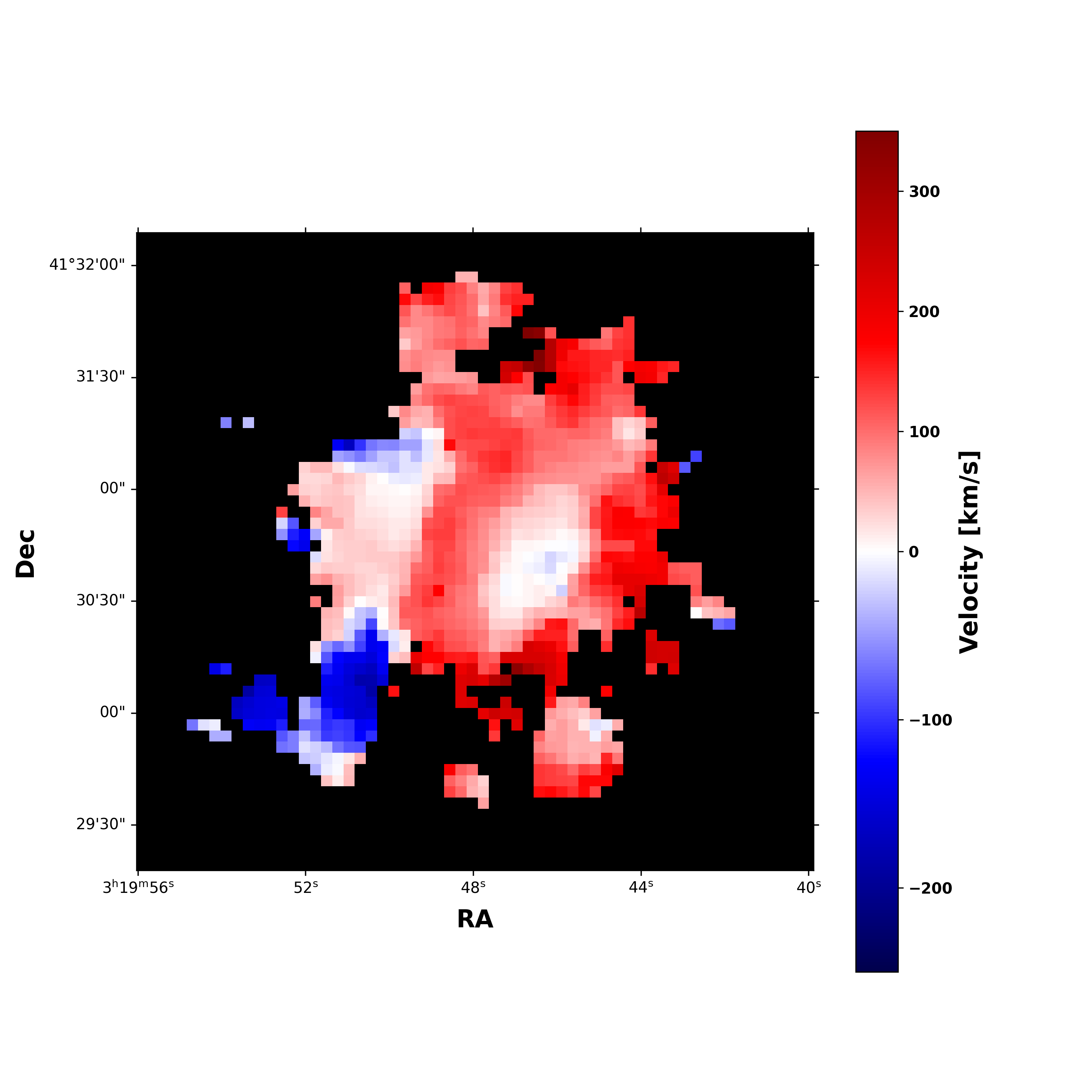}  
    %\label{fig:sub-first}
\end{subfigure}%
\begin{subfigure}{0.5\textwidth}
    \centering
    % include second image
    \includegraphics[width=1.0\linewidth]{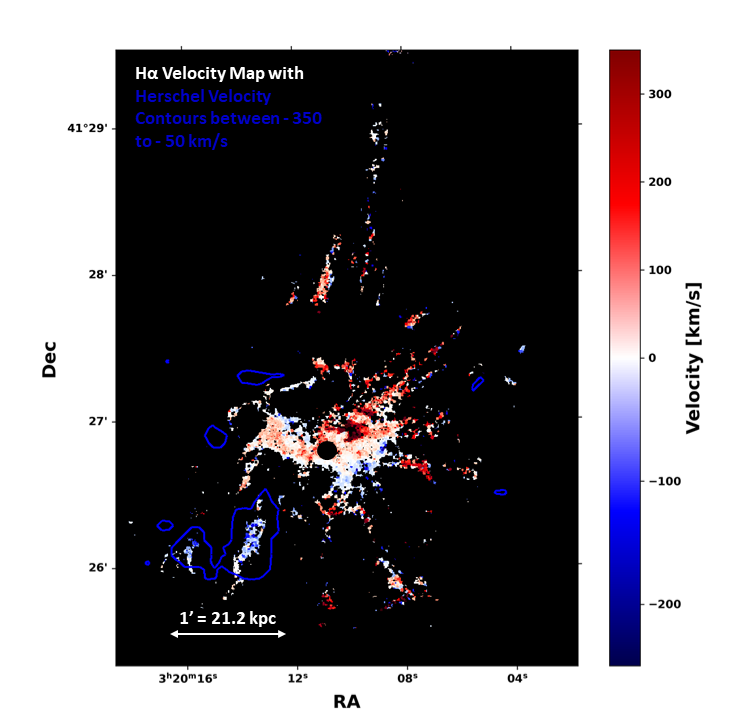}  
    %\label{fig:sub-second}
\end{subfigure}
\newline
\begin{subfigure}{0.5\textwidth}
    \centering
    % include third image
    \includegraphics[width=1.0\linewidth]{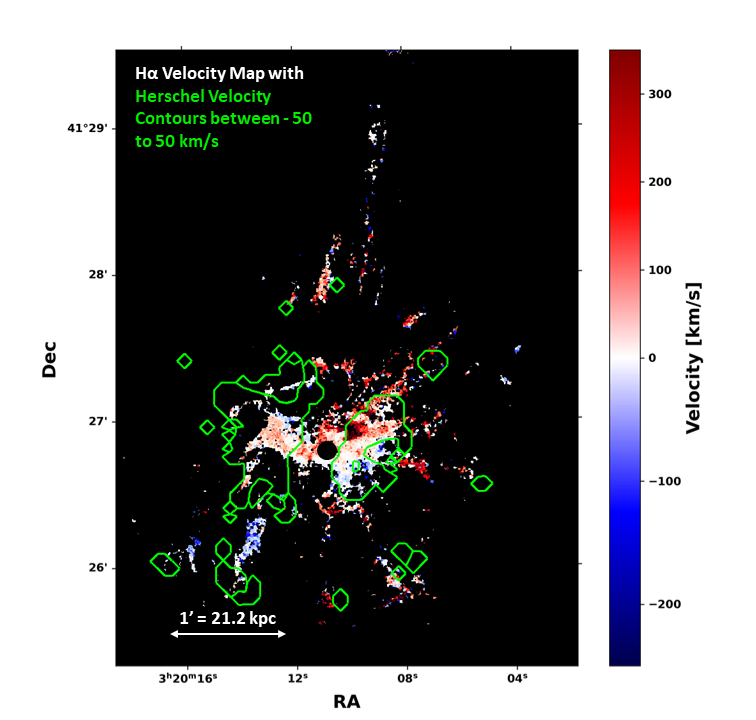}
    %\label{fig:sub-third}
\end{subfigure}%
\begin{subfigure}{0.5\textwidth}
    \centering
    % include fourth image
    \includegraphics[width=1.0\linewidth]{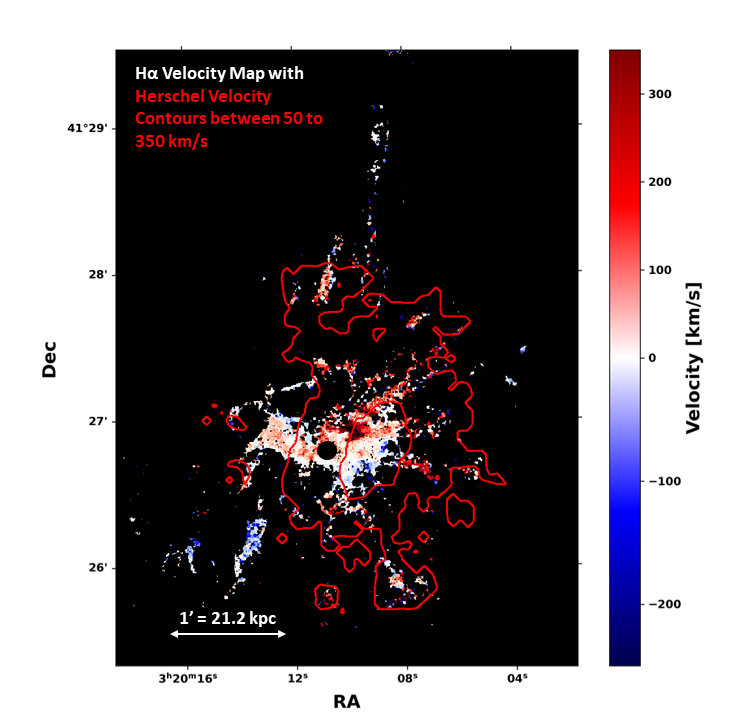}  
    %\label{fig:sub-fourth}
\end{subfigure}
\caption{Upper left: Herschel Infrared Space Observatory [CII] emission line velocity map corrected for a redshift of z = 0.017284 (\citealt{mittal2012herschel}). Upper right: H$\alpha$ velocity map as obtained with SITELLE data at high spectral resolution overplotted with contours of Herschel velocity between -350 to -50 km/s. Bottom left: H$\alpha$ velocity map as obtained with SITELLE data at high spectral resolution overplotted with contours of Herschel velocity between -50 to 50 km/s. Bottom right: H$\alpha$ velocity map as obtained with SITELLE data at high spectral resolution overplotted with contours of Herschel velocity between 50 to 350 km/s.}
\label{fig:herschel_contours}
\end{figure*}

%%% Include the comparison with Mittal et al. to show comoving structures.

One model argues that the depletion of metals in the galaxy and the deformed morphology of these filaments are explained by the rise of bubbles fueled by the activity of the central supermassive black hole (e.g. \citealt{mcnamara_heating_2007}).

A second model demonstrates that the rising of radio bubbles through the ICM can trigger local thermal instabilities in the X-ray emitting gas that would lead to local cooling flows where the density of material would increase as a consequence of cooling.
%thus creating shocks and turbulence, could develop local thermal instabilities in the X-ray emitting gas. 
%\color{red} These would be at the source of local thermal instabilities where the density of material would increase as a consequence of cooling. \color{black}
%Such regions would be at the source of small cooling flows where the density of material would increase as a consequence of cooling. 
Due to the intense gravitational potential of the BCG, these filaments of cooler gas would fall onto the BCG, fueling the black hole activity through this feedback cycle (e.g. \citealt{voit_global_2017}, \citealt{tremblay_galaxy-scale_2018}).
This precipitation-based model would explain the observed multiphase structure by the cooling of gas from X-ray temperatures at $\sim 4 \times 10^7$K, down to cold molecular gas at $< 10^3$K visible in radio (\citealt{mcnamara_mechanism_2016}, \citealt{dutta_cooling_2022}). Thus, the detection of a cold CO(2-1) molecular gas structure close to the central region of the filaments in NGC 1275, as well as similar detections of correlated cold molecular gas in the outer filaments (\citealt{salome_very_2011}) offer a key argument regarding this formation model. The strongly emitting central structure was detected and studied by \cite{lim_radially_2008} and is divided into three main filaments and smaller clumps, all located near the central plane of the galaxy. They also showed that this central disk spatially correlates with the optical filaments seen in H$\alpha$, as well as the X-ray emission between 0.5 and 2.0 keV. The X-ray image shown in Figure \ref{fig:spatial_correlation} was obtained by combining several Chandra observations between 0.5 and 2.0 keV and correcting the exposure of the background subtracted image. The exposure time is comprised between 800 ks and 1 Ms. Our new observations with SITELLE as seen in Figure \ref{fig:spatial_correlation} also show that the H$\alpha$ flux contours spatially correlate with the CO(2-1) molecular gas as seen by \cite{lim_radially_2008}, especially in the case of the left-side filaments of the central disk-shaped structure (see right panel of Figure \ref{fig:spatial_correlation}), while also located along the lower borders of an expanding radio-filled bubble as observed by \cite{gendron-marsolais_high-resolution_2020}. This region seems to be directly linked with the central disk that has higher averaged flux and velocity dispersion as seen with the H$\alpha$ flux map shown in Figure \ref{fig:maps}.

Such contrast between central and extended flux
%This result 
is similar to what was previously obtained by \cite{salome_very_2011}, where the detection of CO across the filaments demonstrated that the central region has a flux around 10 times higher than the emission found in the filaments. However, \cite{lim_radially_2008} clearly showed that this higher central flux emission is not only linked to the AGN, but also to this bright and uniform disk-shaped feature. This result proves insightful when considering the observations of BCGs by ALMA (\citealt{russell_driving_2019}) revealing mostly filamentary and disk-shaped structures around these galaxies. Some of these features display rotation (e.g. \citealt{2014MNRAS.437..862H}), however, this doesn't appear to be the case for the central disk-shaped structure of NGC 1275 (see \citealt{lim_radially_2008}), even though, closer observations of the inner regions close to the central galaxy indicates rotation motions of the gas (see \citealt{nagai_alma_2019}, \citealt{2020MNRAS.496.4857R}). 

Decades of optical observations have revealed that NGC 1275 is surrounded by an extended array of filaments spreading throughout the inner 100 kpc and are the largest seen in any cluster (e.g. \citealt{conselice2001nature}). However, we also know that the Perseus cluster is one of the closest to us, therefore, in more distant clusters, only the brightest filaments could potentially be seen. Indeed, when considering our renewed observations of NGC 1275 with SITELLE, we can observe a significantly brighter central disk-shaped structure ($\sim 1\times10^{-17} \text{ erg s}^{-1}\text{cm}^{-2}\text{\AA}^{-1}$ flux) as opposed to the outer filaments ($\sim 1\times10^{-18} \text{ erg s}^{-1}\text{cm}^{-2}\text{\AA}^{-1}$ flux). Thus, more distant clusters could also harbour extended array of filaments that our current generation of telescopes are not able to detect. Simulations exploring this idea have been developped for NGC 1275 by \cite{2010ApJ...721.1262M} and revealed that by a redshift of $z\sim 0
.06$, most of the filaments would be undetectable leaving only the central bright region surrounded by unresolved features (see Fig. 16 of \citealt{2010ApJ...721.1262M}). Similarly, recent observations of the BCG of the Centaurus cluster by \cite{2016MNRAS.461..922F} revealed a faint and diffuse H$\alpha$ nebula surrounding it.

Recent ALMA observations of NGC 1275 by \cite{nagai_alma_2019} provided a detection of a smaller portion of the central disk in CO(2-1) close to the AGN, as well as HCN(3-2) and $\text{HCO}^+$(3-2) emission within a radius of 1.8 arcsec (0.658 kpc) around the AGN, revealing a rotating motion of the emitting gas at this scale. Such detection is within the central region that we have masked in our SITELLE fits due to the broad component being present as mentionned in Section 2.6. 
% Thus, a kinematical study of the innermost regions could be carried out in a future work after the disentanglement of the AGN and filament emission seen with SITELLE at a similar scale to the observations made by \cite{nagai_alma_2019}. This could enable us to see if a rotating motion is also detected in the optical domain.

% Furthermore, one interesting feature of this central disk detected in radio is its spatial correlation with X-ray and H$\alpha$/optical emission. Indeed, as seen with Figure \ref{fig:spatial_correlation}, there is a definitive link between all these emission structures that would argue for a local cooling region, that could be part of an infalling cooling flow. However, we also notice that this region seems to be located along the lower border of an expanding radio-filled bubble, which is especially visible for the central western filament. This means that the turbulence induced by its expansion (see simulations by Karen Yang et al. 2019) could be the source of the local intense cooling taking place in this region. Other arguments supporting these observations come from the velocity and velocity dispersion maps obtained for H$\alpha$ that will be described in the following subsections.

\subsection{Velocity Dispersion Structure}

%%% REPHRASE that the velocity dispersion in the central region is not increased overall by multiple velocity components even if some knots can be found within this region.

Previous observations of the filamentary nebula surrounding NGC 1275 with slit spectroscopy or IFU instruments demonstrated that the velocity dispersion of the ionised optical gas was higher in the central region near the AGN ($\sim 150$ km/s) while diminishing gradually to lower values ($\sim 50$ km/s) in the outer filaments (\citealt{hatch_origin_2006}, \citealt{gendron-marsolais_revealing_2018}). These observations support the idea that the AGN activity was the source of a higher level of agitation in the central region while other mechanisms such as the turbulence and shocks induced by the trailing of ascending radio-filled bubbles would induce a lower but non-negligible velocity dispersion in the outskirts of the structure (\citealt{falceta-goncalves_turbulence_2010}). 
%The velocity dispersion map produced by \cite{gendron-marsolais_revealing_2018} with SITELLE supported this model by showing similar values.

\begin{figure*}
\begin{subfigure}{0.5\textwidth}
    \centering
    \includegraphics[width=1.0\linewidth]{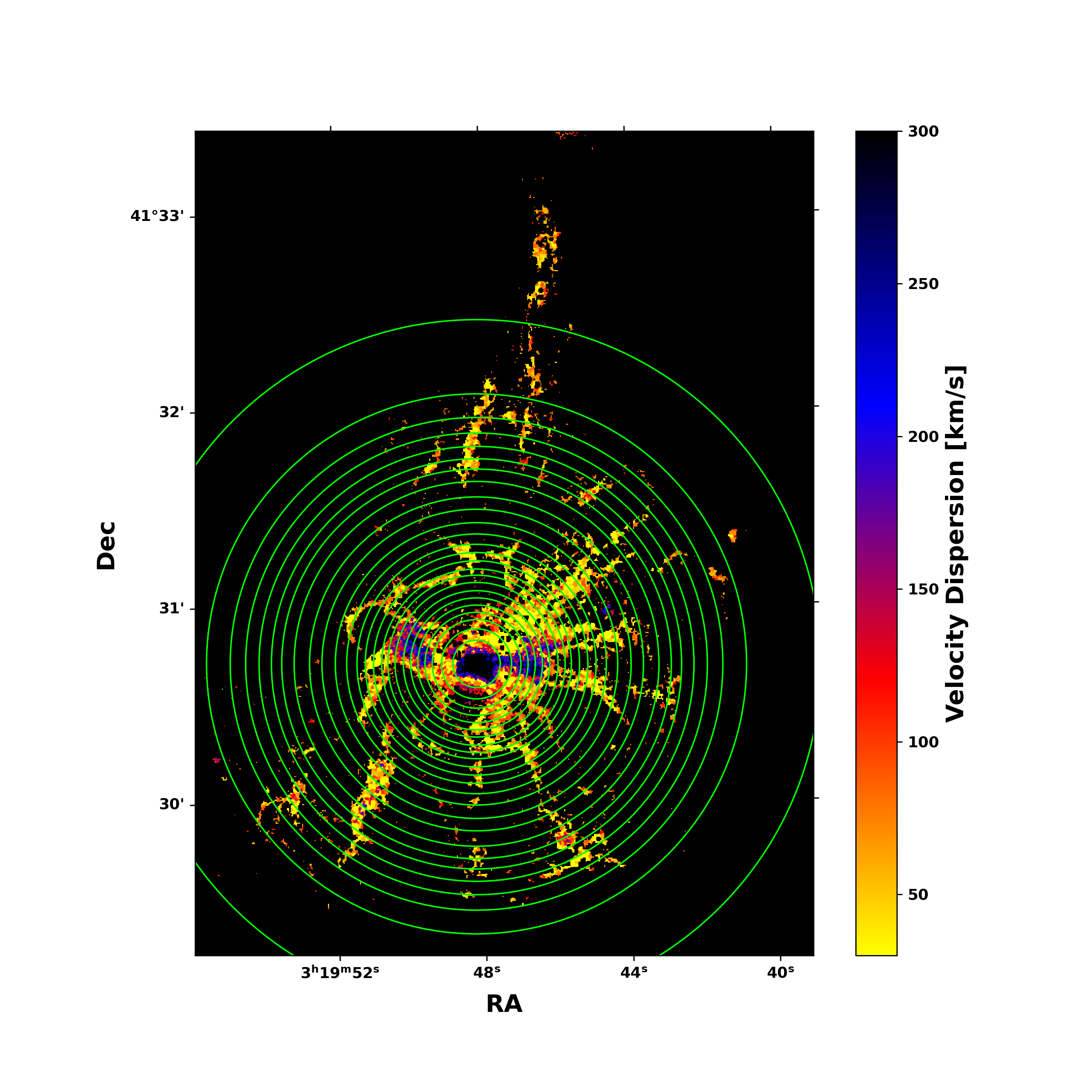}  
\end{subfigure}%
\begin{subfigure}{0.5\textwidth}
    \centering
    \includegraphics[width=0.95\linewidth]{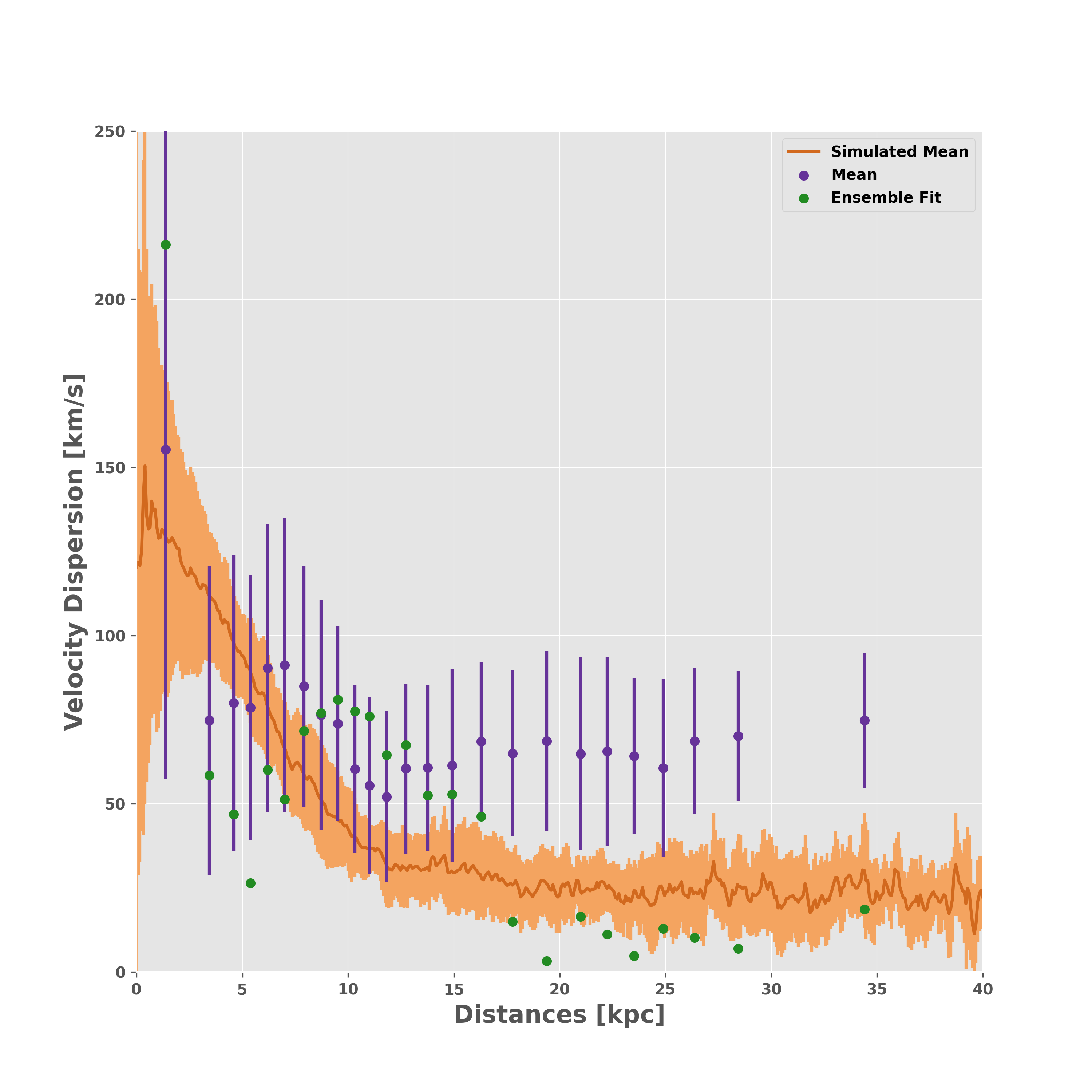}  
\end{subfigure}
\caption{Left: H$\alpha$ velocity dispersion map. Right: Mean (in purple) and ensemble fits (in green) across annuli containing 1500 pixels for the H$\alpha$ velocity dispersion. The error bars for the ensemble fit have been plotted but are too small to see compared to the data scale. Simulation results of the velocity dispersion in optical filaments of the Perseus Cluster. The brown line shows the average line-of-sight velocity dispersion as a function of radius in the simulated Perseus cluster from \cite{li2015cooling}, while the 1$\sigma$ scatter is shown though the orange shaded area.}
\label{fig:velocity_dispersion}
\end{figure*}

The new SITELLE observations at high-spectral resolution confirm that the central disk-shaped structure with higher velocity dispersion is not due to multiple velocity components based on visual inspection of the spectra. Indeed, the regions displaying multiple velocity components are extremely localized compared to the extended disk-shaped central structure visible in Fig. \ref{fig:maps}. 
Beyond this disk-shaped structure, the velocity dispersion decreases sharply and remains uniformly low throughout the rest of the filaments, out to $r \sim 50$ kpc. The central velocity dispersion is almost two times higher in the central disk-shaped structure than what previous low-spectral observations showed (\citealt{gendron-marsolais_revealing_2018}, see left plot of Fig. \ref{fig:velocity_dispersion}). When considering only the central bright region by applying a more stringent flux cut-off of $1\times10^{-17} \text{ erg s}^{-1}\text{cm}^{-2}\text{\AA}^{-1}$, the mean velocity dispersion of the central disk-shaped structure is $\sim 134$ km/s. The disk-shaped structure seems specifically located at the position of the central CO(2-1) disk of molecular gas and at the border of the inflating radio bubble (see right side of Figure \ref{fig:spatial_correlation}).
%, as can be seen in the right side of Figure \ref{fig:spatial_correlation}, where the radio bubble is delimited by the cyan contours as observed in radio by \cite{gendron-marsolais_high-resolution_2020}. 

The most striking feature displayed in Fig. \ref{fig:velocity_dispersion} is the lack of a smooth gradient between the high dispersion of the central region ($\sim 134$ km/s) and the lower dispersion of the outer filaments ($\sim 44$ km/s - see also the right plot of Figure \ref{fig:velocity_dispersion}). For the gas in the central region, we could expect to see a higher velocity dispersion since more gas is likely found near the central regions,
whereas for the outer filaments, we are viewing individual filaments and therefore, we are probing more of a pencil beam view of the kinematics in the outer filaments. However, the sharp contrast in velocity dispersion from the central disk-shaped structure to the filaments beyond this structure is puzzling. We also note that despite this expected overlap of gas in the central region, few highly separated velocity component emission lines are found in this region, which will be discussed in section 3.7. 

%%% Mark Voit explanation of pencil beam vs global velocity dispersion

If we consider the entire system of low dispersion filaments, it has a global velocity dispersion much greater than $\sim 44$ km/s, as indicated by the H$\alpha$ velocity dispersion map, even if the central region displaying higher values is excluded. The uniformity in the (pencil beam) velocity dispersion of the outer filaments might therefore be reflecting a different physical process from the one responsible for the global velocity dispersion of the filament system.

The global velocity dispersion is likely to be connected to motions of the hot medium, while the pencil beam velocity dispersion of the ionized gas is more likely to be connected to shearing motions at the interfaces between the filaments and the hot medium, which we will investigate in section 3.5.

Thus, this result could also imply that two completely different mechanisms might be at play to introduce such a clear differentiation in velocity dispersion.

%We also incorporated simulations results of velocity dispersion in optical filaments for the Perseus Cluster with data coming from \cite{li2015cooling}, which can be seen in the right plot of Figure \ref{fig:velocity_dispersion}. 
We also compare our result with simulations from \cite{li2015cooling} (see Figure \ref{fig:velocity_dispersion}). The brown line shows the average line-of-sight velocity dispersion as a function of radius in simulated filaments from \cite{li2015cooling} that reproduce the ones seen in NGC 1275. 
We first take the simulation outputs generated every 10 Myr over a 300 Myr period when the simulated cluster most resembles Perseus in the morphology and spatial distribution of the H$\alpha$ filaments. For each output, we make a line-of-sight velocity dispersion map of the H$\alpha$ gas (with temperatures of $\sim 10^4$ K) weighted by emissivity. Then, the data is split into radial distance bins and averaged for each bin. We then compute the time-averaged velocity dispersion profile and its 1$\sigma$ scatter (orange shaded area). We can thus see that the simulation shows lower values of velocity dispersion further away from the central region which is in accordance with our observational results. 
Overplotted in purple and green are the mean and ensemble fits extracted from the annuli seen in the left side of \ref{fig:velocity_dispersion}. These datapoints clearly show a drop in velocity dispersion at a short radial distance away from the central region and matching the simulated data. The outer filaments display stable low velocity dispersion in agreement with simulated data.
%even if a discrepancy is found between the mean and ensemble fit for the outer filaments. We argue that such differences might be caused by the difference in methodology used to produce the mean and ensemble fits as well as the spatial distance between pixels within an annuli affecting the ensemble spectrum.\color{black}

An interesting similarity to our results can be observed in the study of Abell 2597 by \cite{tremblay_galaxy-scale_2018} as observed with MUSE and ALMA. This study also demonstrated the presence of a spatial correlation between a central v-shaped molecular gas structure and its optical emission counterpart in H$\alpha$. Both of these structures were also found to be comoving and closely tracing the wake of the X-ray cavities formed by the activity of the BCG's AGN. Moreover, when considering the velocity dispersion of the optical gas in Abell 2597 (see Figure 13 of \citealt{tremblay_galaxy-scale_2018}), a similar result can be observed where the correlated emission region displays an extremely high velocity dispersion (up to $\sim 350$ km/s) which quickly falls to lower values for the outer filaments ($\sim 50$ km/s) can be seen. 

%These similar trends reinforce the fact that NGC 1275 and Abell 2597 have several similar properties such as a wake that is found between the H$\alpha$ filamentary nebula and the X-ray cavities (\citealt{tremblay_galaxy-scale_2018}), as well as an interaction between the inflating radio bubbles and the central bright H$\alpha$ and molecular gas structures closely following its edges (see the left side of Fig. \ref{fig:spatial_correlation}). 

Regarding the filament formation models that have been developed and detailed previously, \cite{tremblay_galaxy-scale_2018} explore a unifying model involving chaotic cold accretion, precipitation, and stimulated feedback through the description of a galaxy-scaled "fountain". In this model, the central cold molecular gas and optically emitting filaments would supply the accretion reservoir of the AGN, thus fueling its activity. The jets it produces would then inflate radio bubbles, buoyantly rising and carving the surrounding ICM, therefore creating turbulence and thermal instabilities as well as uplifting ionised gas in their trail (e.g. \citealt{mcnamara_mechanism_2016} and simulation works by \citealt{2014ApJ...789..153L}). The raised material would then precipitates back to the central potential well thus sustaining the "fountain". Moreover, further thermal instabilities generated by the turbulent AGN feedback would reinforce the precipitation of colder gas in slightly denser parts of the ICM, thus facilitating the formation of extended filamentary nebulae (e.g. \citealt{Voit_2018}, see also \citealt{2014ApJ...785...44M} and \citealt{russell_driving_2019}).

% Velocity structure function (VSF) have been used to characterized the underlying dynamics at play in the filamentary nebula \cite{li_direct_2020}. The VSF calculated from our new high-spectral resolution observations will be presented in a future paper.

\subsection{Extended Filaments and Turbulent Radiative Mixing Layers}
%\subsection{\color{blue}Turbulent Radiative Mixing Layers and Hidden Cooling Flows\color{black}}

%%% ADD introduction on Turbulent Radiative Mixing Layers 

We now investigate a possible formation scenario for the extended filaments displaying lower velocity dispersion through turbulent radiative mixing layers. This scenario involves the formation of filaments in situ instead of beeing dragged out by the adiabatic rise of the radio bubbles carved by the SMBH jets.

Turbulent mixing layers are frontier layers between gaseous materials at different temperatures. Due to their extreme difference in internal energy, the materials will interact and produce a layer of intermediate temperature where turbulent motions are more prominent. This layer will thus produce emission lines associated with the intermediate temperature and possibly shield the inner colder layer from being completely mixed with the higher temperature material. In the context of filaments in clusters, the hot intracluster gas at $\sim 4 \times 10^7$ K (\citealt{hitomi_collaboration_quiescent_2016}) acts as the warm layer while turbulent motions are thought to be due to the AGN feedback processes and set a frontier layer between this medium and the colder optically emitting filaments at $10^4$ K.

Previous works by \cite{begelman1990turbulent} have highlighted the idea of extended turbulent mixing layers in the intracluster medium. They discussed that filamentary structures visible in the optical could form out of the cooling flows of AGN through absorption of far UV emission of turbulent mixing layers, which would then induce UV and optical emission by photoionization of the colder gas. However, detections of embedded cold molecular gas all throughout the filamentary nebula were made later by \cite{salome_cold_2006}, implying that turbulent radiative mixing layers could be created at the boundaries between cold molecular gas and the surrounding hot ICM.
%Therefore, the presence of colder gas within the filaments would imply that for a turbulent mixing layer to form, its temperature should be lower and thus closer to optically emitting temperatures. %%% A REFORMULER

Indeed, the ICM of the Perseus cluster is known to have a high temperature ($\sim 4\times10^7$ K, see \citealt{hitomi_collaboration_quiescent_2016}), while the cold molecular gas found all throughout the filaments has a much lower temperature ($\sim 10-10^3$ K, e.g. \citealt{salome_very_2011}). Thus, we might wonder if the optical emission observed with the SITELLE data could be part of a radiative turbulent mixing layer between the two media. The presence of a turbulent mixing layer could then explain the intermediate temperature of the optical gas.

%%% DISCUSS here paper on hidden cooling flows.

 We do know that the filaments are extremely thin and display a thread-like structure (\citealt{fabian_magnetic_2008, fabian2016hst}) which would be reminiscent of the shape taken by turbulent mixing layers (see \citealt{2022arXiv221015679A}). 
 %This could imply that extended filaments could be part of a turbulent mixing layer between the hot ICM and the cold molecular gas. 
 %With such a thin structure, multiple emission line components for the outer filaments are difficult to perceive which could mean that physically close filaments would be connected thus leading to single component emission lines
With such a thin structure, physically close filaments must be connected, leading to single component emission lines (see \citealt{hatch_origin_2006}). A higher spectral resolution than $R=7000$ would be needed to explore in detail the possible multiple components of their emission lines since none are detected within the outer filaments after a spectral examination of the SITELLE data.
 %with the SITELLE data after a spectral examination of several outer filaments.

The velocity dispersion in mixing layers are increased by in situ turbulence caused by the interaction between material at different temperatures. Recent simulations (e.g. \citealt{ji2019simulations}, \citealt{2022ApJ...924...82F}, \citealt{2022arXiv221015679A}) show however that the turbulence of material inside the layer remains low ($\sim 30$ km/s.). These simulations results could potentially prove insightful in regards of the observed mean velocity dispersion value of $\sim 44$ km/s, prevalent within the extended filamentary nebula surrounding NGC 1275.

Hence, when considering the structure of a turbulent mixing layer in conjunction with our optically emitting extended filaments, it could mean that the outer layer of turbulence between the cold molecular gas and the hot ICM could be shielding the inner filaments and preventing them from having a higher velocity dispersion, leading to a clear dichotomy between inner filaments, disturbed by the formation and growth of radio bubbles, and outer quiescent filaments. Therefore, the velocity dispersion in outer extended filaments could be a proxy for turbulent radiative mixing layers.

Finally, recent works reported the discovery of hidden cooling flows in a sample of nearby clusters of galaxies, including Perseus (see \citealt{fabian2022hidden}, \citealt{fabian2023ahidden}, \citealt{fabian2023bhidden}). Here, the authors argue that AGN feedback may not be as efficient in heating the intracluster medium as initially thought - instead, the intracluster medium would be allowed to cool, explaining the substantial molecular gas reservoirs found in these systems. They predict that these cooling flows are not seen at X-ray wavelengths because they are being obscured by photoelectrically-absorbing cold clouds and dust (e.g. \citealt{white1991discovery}, \citealt{johnstone1992spectral}, \citealt{fabian1994asca}) - if this is the case, then the absorbed emission will re-emerge in far infrared. 

Indeed, extreme cooling of the gaseous material surrounding BCGs could prohibit them from being detected due to extremely low temperature and emissivity (see \citealt{fabian2023ahidden}). According to these results cooling flows of $\sim 30 - 100$ $M_{\odot}\text{.yr}^{-1}$ could therefore be present within clusters such as the Perseus cluster of galaxies (\citealt{fabian2022hidden}). Interestingly, it is suggested that these hidden cooling flows could have internal structures resembling that of turbulent radiative mixing layers, similar to those proposed here (see \citealt{begelman1990turbulent}, \citealt{fabian2023ahidden}). Nevertheless, future observations of the ICM in clusters of galaxies with upcoming X-ray telescopes such as XRISM will probe a variety of transitions expected from such hidden cooling flows, providing further insight into these flows and their connection to multiphase gas in clusters.

%Finally, recent works on the subject of cooling within filamentary nebulae and the multiphase gas surrounding the BCGs of cool-core clusters demonstrate that hidden cooling flows from the X-ray spectra of BCGs could be accounted for in the far infrared (see \citealt{fabian2022hidden}, \citealt{fabian2023ahidden}, \citealt{fabian2023bhidden}). Indeed, the apparent lack of X-ray emission from these sources was thought to be the result of an absorption phenomenon (e.g. \citealt{white1991discovery}, \citealt{johnstone1992spectral}, \citealt{fabian1994asca}). 

%Talk about the hidden cooling flows model linked with the presence of cold molecular gas embedded within the filaments. Missing X-ray emission associated with photoelectric absorption.

% Change conclusion accordingly to mention the model and the hidden coling flows.

\subsection{Velocity Structure}

\begin{figure*}
\begin{subfigure}{0.5\textwidth}
    \centering
    \includegraphics[width=1.0\linewidth]{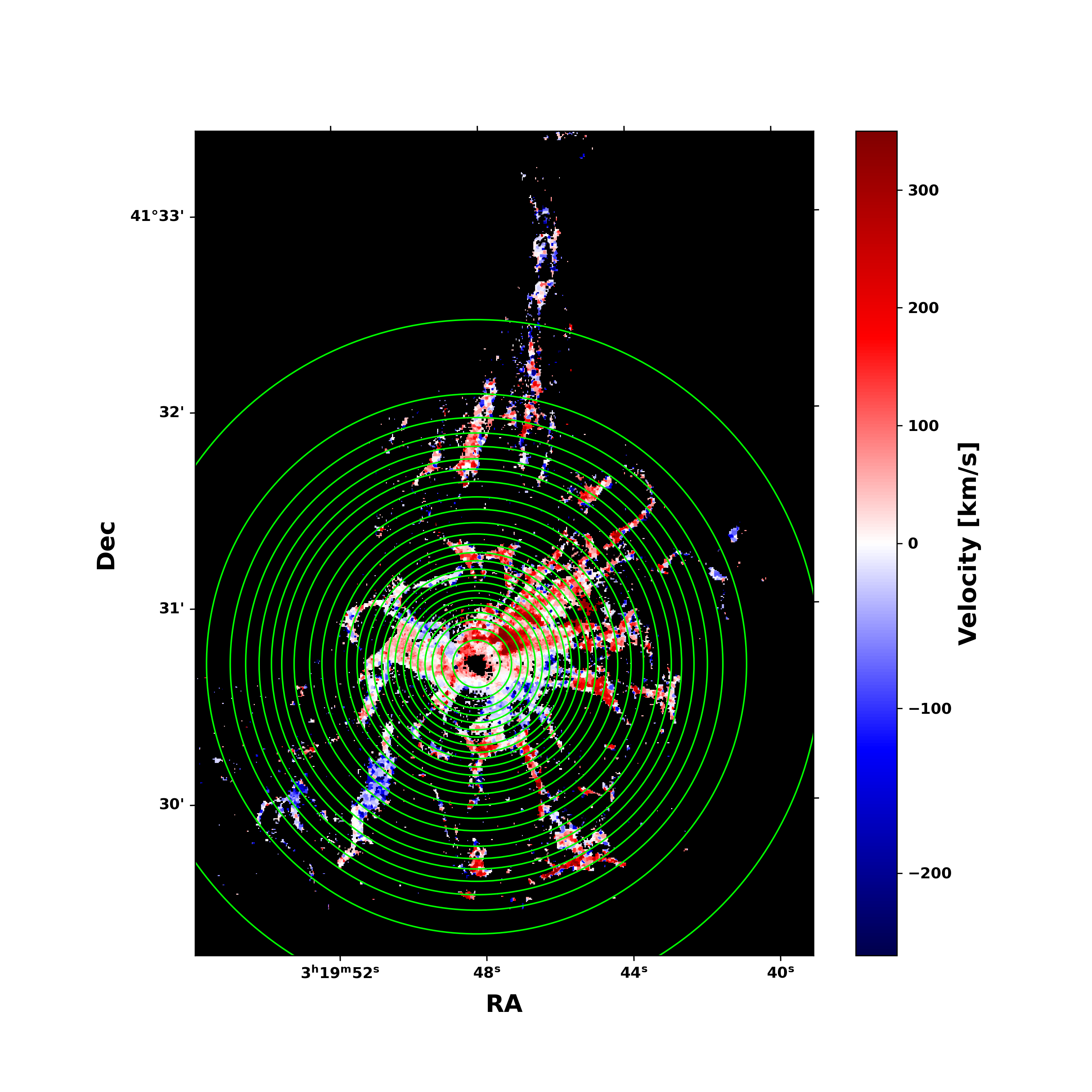}  
\end{subfigure}%
\begin{subfigure}{0.5\textwidth}
    \centering
    \includegraphics[width=0.95\linewidth]{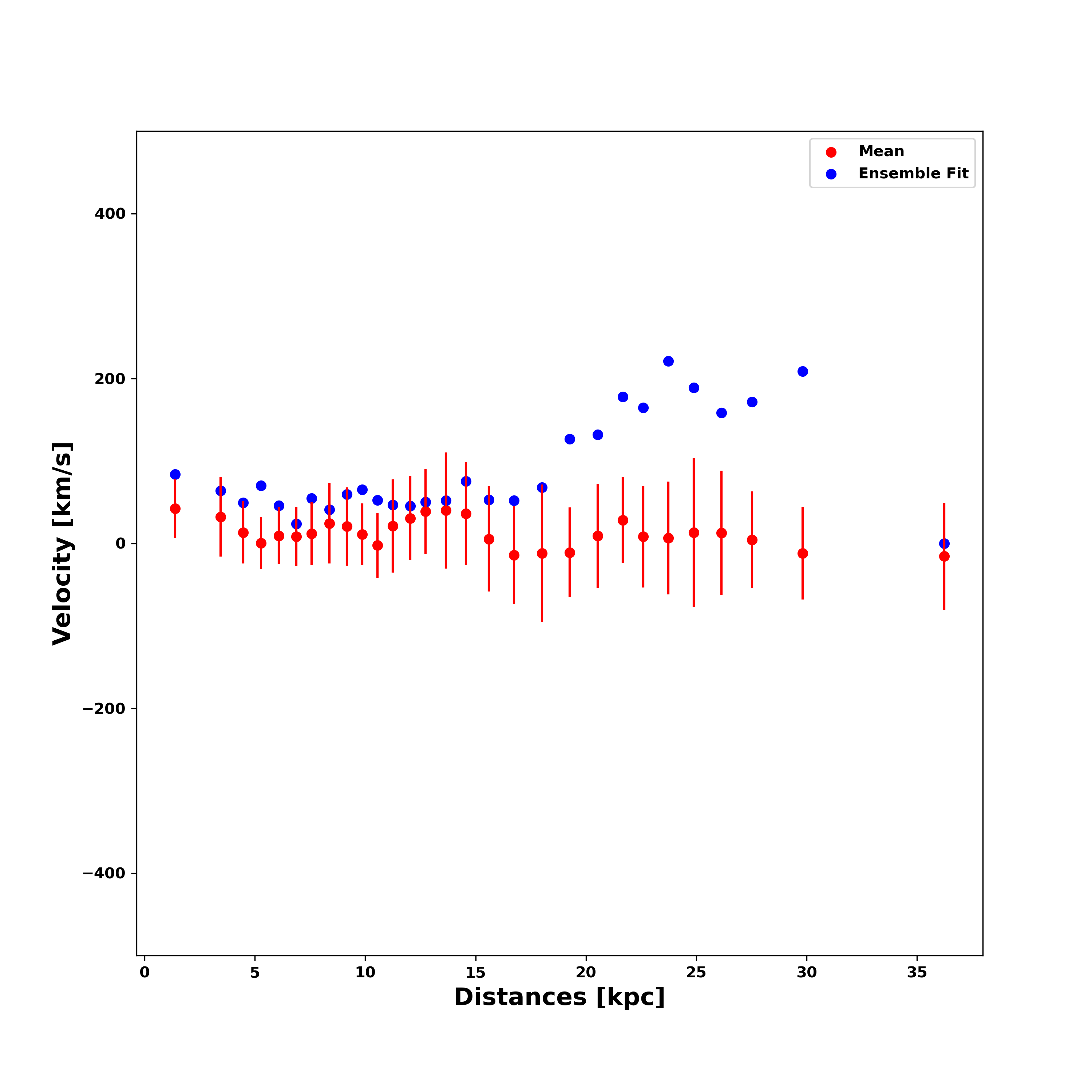}  
\end{subfigure}
\caption{Left: H$\alpha$ velocity map. Right: Mean (in red) and ensemble fits (in blue) across annuli containing 1500 pixels for the H$\alpha$ velocity. The error bars for the ensemble fit have been plotted but are too small to see compared to the data scale.}
\label{fig:velocity}
\end{figure*}

%%% USE symetrical colorbar ! Mention which region is presented.

Now focusing on the analysis of the velocity structure obtained with the new SITELLE observations, we find a very similar H$\alpha$ velocity map as to what was previously obtained by \cite{gendron-marsolais_revealing_2018} at lower spectral resolution observations (see Figure 5 of \citealt{gendron-marsolais_revealing_2018} and left plot of Figure \ref{fig:velocity}). 
This result strengthens previous conclusions regarding the dynamics of the filamentary nebula. 
Moreover, when looking at the mean and ensemble velocity fits across annuli containing 1500 pixels (see right plot of Figure \ref{fig:velocity}), there does not appear to be a specific radial velocity trend throughout the filaments, which reinforces the idea of a chaotic velocity structure.

Considering the Herschel infrared emission kinematics as studied by \cite{mittal2012herschel} and displayed in the upper left panel of Figure \ref{fig:herschel_contours}, we can see that similar velocity trends could also hint at a comoving nature of the gas in various wavelengths. Indeed, when considering similar velocity intervals as Figure 5 in \cite{mittal2012herschel}, we can clearly see that the south-western filaments kinematics (see top-right plot of Figure \ref{fig:herschel_contours}) display a similar trend of negative velocities between $-350$ and $-50$ km/s with the spatially correlated [CII] emission line kinematics (see top-left plot of Figure \ref{fig:herschel_contours}). This trend is also visible for both intermediate velocity values between $-50$ and $50$ km/s as well as positive velocities between $50$ and $350$ km/s, although less strikingly (see bottom plots of Figure \ref{fig:herschel_contours}).

\begin{figure}[h]
    \centering
    \includegraphics[width=95mm,scale=0.5]{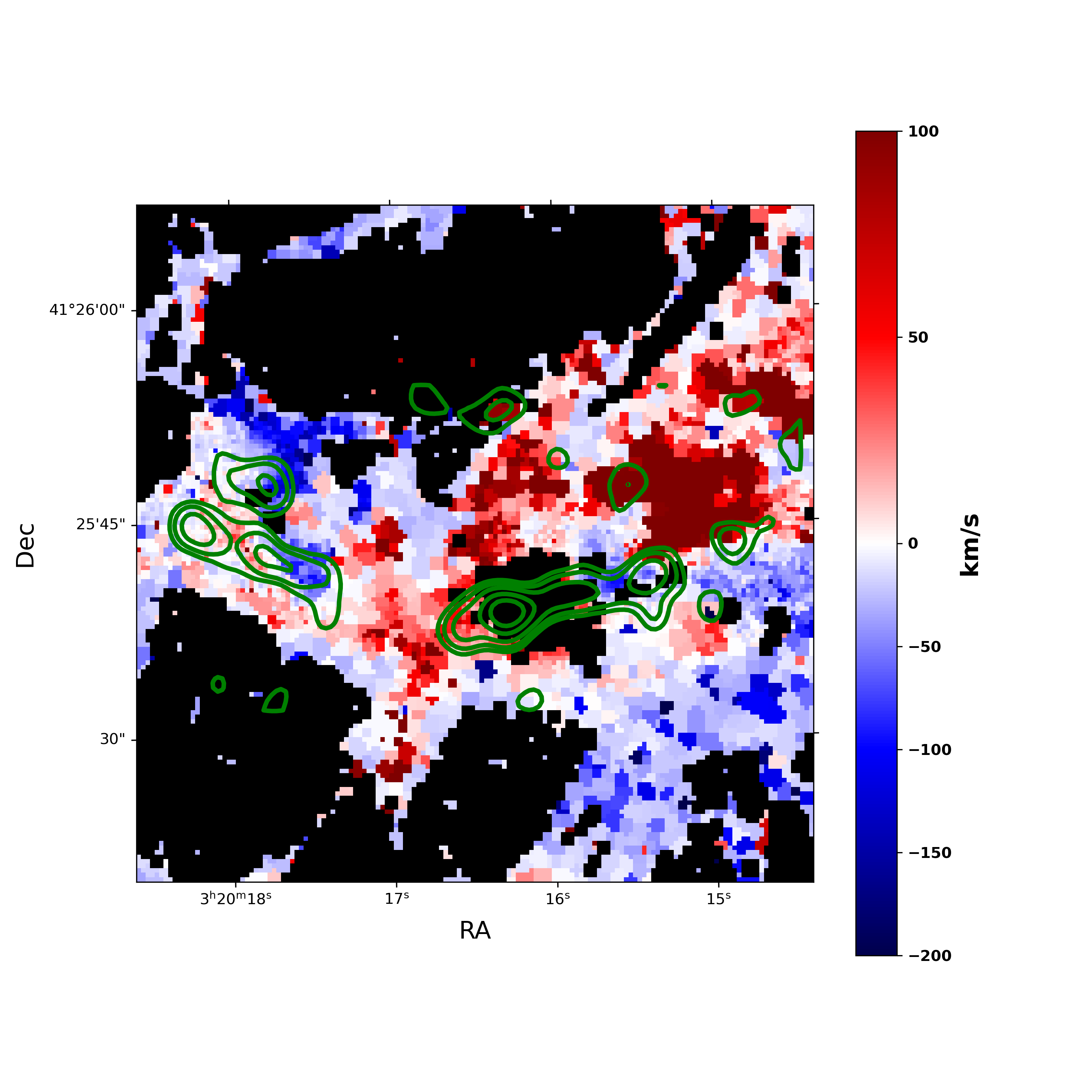}
    \caption{Velocity map in H$\alpha$ and \nii\text{ } of the central region of the filamentary nebula shifted to a redshift of z=0.01756 in order to match the study of \cite{lim_radially_2008}. The contours of the central disk observed in CO(2-1) with the SMA by \cite{lim_radially_2008} are superimposed in green.}
    \label{fig:velocity_central_disk}
\end{figure}

\cite{lim_radially_2008} also obtained a velocity map of the central molecular gas disk, which showed mostly blue-shifted velocities and without sign of global rotation. To compare directly the velocity maps obtained for the optical emission to the molecular map, we shifted our map considering a redshift value of $z = 0.01756$ used by \cite{lim_radially_2008} instead of the adopted $z = 0.017284$. After this correction, the H$\alpha$ emission shows a similar velocity structure for the central disk-shaped structure (see Figure \ref{fig:velocity_central_disk} and Figure 2 of \citealt{lim_radially_2008}).

\begin{figure*}
    \centering
    \includegraphics[width=180mm,scale=0.5]{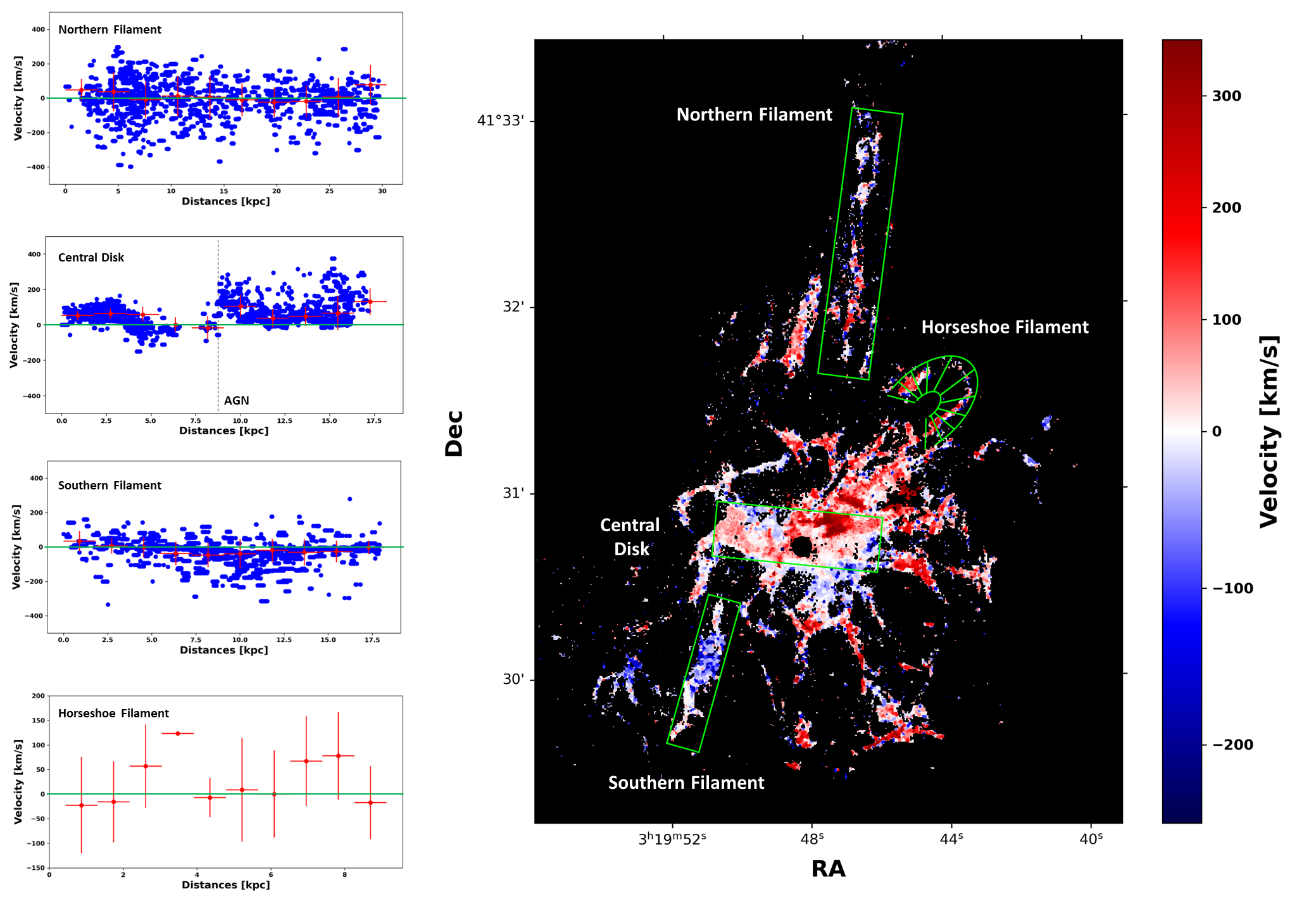}
    \caption{Velocity profiles of selected filaments in the nebula surrounding NGC 1275 (in km/s with the green lines showing a null velocity). For the first three profiles, we show in blue the velocity of each pixel bin found in the region, while the red points illustrates the mean taken in ten subregions of equal widths. The velocity profile of the Horseshoe filament only displays the mean velocity across 10 subregions.}
    \label{fig:velocity_profiles}
\end{figure*}

The detection of similar disk-shaped structures can also be found in simulation works such as those done by \cite{li_modeling_2014}, where they exhibit higher levels of flux and originate from the cooling of gas onto the black hole's potential well. However, the main difference to these simulations is the fact that the central disk always displayed a rotating motion along the plane perpendicular to the jets, which is not the case for our observations where the motion seems more chaotic (see Figures \ref{fig:velocity} \& \ref{fig:velocity_central_disk}). This difference shows that these simulations seems to not be fully capturing  the physical processes at play in the formation and evolution of the filaments we are observing in NGC 1275. Moreover, the presence of radio bubbles fueled by the SMBH relativistic jets, which seem spatially linked to the presence of this disk, was not considered in previous simulations until recently (see \citealt{yang_impact_2019}). However, these simulations incorporating the effect of expanding radio bubbles do not yet study their impact on the velocity dispersion and kinematics of the surrounding medium of the bubbles. This aspect of 3D hydrodynamical simulations of AGN feedback still needs to be explored to properly understand the prominent role of radio bubbles on their surrounding environment.

%%% MENTION if Yang simulations match the observed results we have.

In that sense, recent simulation works by \cite{2022MNRAS.517..616Z} clearly show the turbulence and eddy formation as well as the uplift of material in the wake of rising radio bubbles. Though the simulation models only involve detached bubbles approaching their terminal velocities, it is still of interest to see that material close to the central region gets lifted upwards and acts as the upper layer of the deformation close to the bubble's lower boundary (see top panels of Figure 4 from \citealt{2022MNRAS.517..616Z}). Therefore, an argument could be made that the central region or disk visible in our observations, displaying a higher level of velocity dispersion as well as a clear spatial correlation to cold molecular gas, could be lifted through the detachment and adiabatic rise of the bubbles thereafter forming future filamentary structures around NGC 1275.

Thanks to the high-spatial and spectral resolution of SITELLE, we can also study the velocity profiles of targeted filaments in their entirety. To do this, we extracted the velocity measurements and means within 10 bins over some specific filaments as in \cite{gendron-marsolais_revealing_2018}. Figure \ref{fig:velocity_profiles} shows that the velocities of the outer filaments are significantly more heterogeneous and chaotic, while the velocities of the central disk display less variations. Its velocity profile could also hint at a potential rotation pattern in the central radius of 2.5 kpc around the AGN, however, the error bars prevent us from making a clear conclusion.

The northern filament displays a velocity profile ranging from slightly positive to slightly negative values the further away from the base of the filament, which would indicate stretching. Similar results were obtained both with SITELLE (\citealt{gendron-marsolais_revealing_2018}) and the Gemini Multi-Object Spectrograph (\citealt{hatch_origin_2006}). The southern filament mostly shows negative velocities slightly decreasing then increasing along the filament, while the central filament seems to display an inverse trend with mostly positive velocity values. If we now bring our attention to the averaged velocity profile of the Horseshoe filament, we can see that both its bases (first and last datapoints of the lower left panel of Figure \ref{fig:velocity_profiles})  seem to display an average velocity closer to negative velocities of $\sim - 25$ km/s while the tip of the Horseshoe (represented by the fifth and sixth datapoints of the lower left panel of Figure \ref{fig:velocity_profiles}) also display a drop in averaged velocity closer to a null velocity. On the other hand, the branches of the Horsehoe clearly display positively increasing then decreasing averaged velocity values (as shown in the lower left panel of Figure \ref{fig:velocity_profiles}). This averaged velocity profile could indicate that the branches of the Horsehoe filament might be stretching horizontally as opposed to the bases and tip who seem to follow a trend of displacement through the traction of a radio emitting bubble (see \citealt{hatch_origin_2006}). However, the limited detection of the tip of the Horseshoe, as can be seen on the right-hand side of Figure 16, prevents us from making affirmative statements regarding these central bins of the Horseshoe averaged velocity profile.

\section{Conclusion} \label{sec:highlight}

%%% MENTION if the precipitation limit hypothesis is confirmed with this target.

We performed the analysis of new high-spectral resolution observations with SITELLE and obtained new flux, velocity, and velocity dispersion maps of \sii$\lambda6716$, \sii$\lambda6731$, \nii$\lambda6584$, H$\alpha$(6563\AA), and \nii$\lambda6548$ emission lines for the filamentary nebula surrounding NGC 1275, the brightest cluster galaxy of the Perseus cluster.
\begin{itemize}
    \item We detected a central disk-shaped structure displaying higher averaged flux ($\sim 1 \times 10^{-17} - 2 \times 10^{-16}$ erg.s$^{-1}$.cm$^{-2}$.$\AA^{-1}$) and velocity dispersion ($\sim 134$ km/s) than the rest of the filaments which is also spatially correlated with a disk-shaped structure as seen in CO(2-1) associated with molecular gas. Both of these structures seem to spatially correlate with the wake of the radio bubbles that have been inflated through the relativistic jets of the supermassive black hole at the center of the galaxy. However, this disk-shaped feature does not display a clear rotation pattern which entails a definitive difference from similar structures obtained through simulations.
    \item The rest of the filamentary structure displays fainter flux measurement ($\sim 1 \times 10^{-18}$ erg.s$^{-1}$.cm$^{-2}$.$\AA^{-1}$) as well as a much lower velocity dispersion ($\sim 44$ km/s) across the outer filaments, thus implying a potential more quiescent formation mechanism.
    % \item We also obtained new line ratio maps across the entirety of the filamentary nebula. Moreover, we managed to produce the first complete map of the \sii$\text{ }$ emission ratio, giving us new information on the density of the gas that is comparable to previous measurements done in X-ray.
    \item Thanks to our very high-spectral resolution, we also managed to detect regions and knots displaying multiple emission line components. However, they are extremely localized and only found in the central part of the filamentary nebula within $r<10$ kpc. Nevertheless, a future study of these structures could potentially help us to gain more insight into the three-dimensional structure of the central filaments.
    
    %\color{blue}\item We also explored a model of radiative turbulent mixing layers to potentially explain the averaged velocity dispersion properties of the outer filaments of the filamentary nebula. Similar structures could also be involved to explain the hidden cooling flows (see \citealt{fabian2022hidden}) that could be detected within the surrounding environment of NGC 1275 thus explaining the apparent lack of X-ray emission within the AGN spectra.\color{black} 
    
    \item Regarding the formation model explored for this filamentary nebula, it seems that a unifying model as explored by \cite{tremblay_galaxy-scale_2018} could potentially explain the observed characteristics of the emitting gas observed flux and velocity dispersion measurements obtained through the analysis of the SITELLE observations of NGC 1275. Indeed, regarding the galaxy-scale fountain model explored by \cite{tremblay_galaxy-scale_2018} we observe similar trends both in terms of kinematics and velocity dispersion between NGC 1275 and Abell 2597. These similarities are also seen in the comoving nature of filamentary nebulaes with both cold molecular structures and surrounding gas. Therefore, a unifying model of formation through the inflow of cold molecular clouds toward the AGN could thus feed its accretion and drive the formation of radio bubbles. This would in turn uplift multiphase material away from the BCG which could then precipitate back as toward the AGN and supply the fountain, while turbulent radiative mixing layers could form between the hot ICM and cold molecular gas. However, contentious points still remain to be explored to properly comprehend the mechanisms leading to the formation of such extended filamentary nebula.
\end{itemize}

Through our analysis of new high-spectral resolution observations of the filamentary nebula surrounding NGC 1275, we reinforced the previous results established by \cite{gendron-marsolais_revealing_2018} and discovered new structures in the optical emission of the filaments. However, these are the first results obtained with this dataset and we expect to get more through an improved analysis of the emission close to the AGN, a study of the SN1 and SN2 filters of SITELLE, as well as the calculation of the velocity structure function with these new maps. Finally, new X-ray observations with the XRISM space telescope (XRISM Science \citealt{team2020science}), the successor of Hitomi, will enable a breakthrough in the study of AGN feedback. This study will offer a touchstone for the analysis of renewed X-ray observations of NGC 1275.

\acknowledgments

The authors would like to thank the Canada-France-Hawaii Telescope (CFHT) which is operated by the National Research Council (NRC) of Canada, the Institut National des Sciences de l'Univers of the Centre National de la Recherche Scientifique (CNRS) of France, and the University of Hawaii. The observations at the CFHT were performed with care and respect from the summit of Maunakea which is a significant cultural and historic site.

B.V. acknowledges financial support from the physics departement of the Université de Montréal. J.H.-L. acknowledges support from NSERC via the Discovery grant program, as well as the Canada Research Chair program. M.G.-M. acknowledges financial support from the grant CEX2021-001131-S funded by MCIN/AEI/ 10.13039/501100011033, from the coordination of the participation in SKA-SPAIN, funded by the Ministry of Science and Innovation (MCIN). J.L. acknowledges support from the Research Grants Council of Hong Kong through grant 17300620. N.W. is supported by the GACR grant 21-13491X. Y.L. acknowledges financial support from NSF grants AST-2107735 and AST-2219686, NASA grant 80NSSC22K0668, and Chandra X-ray Observatory grant TM3-24005X. L.R.-N. is grateful to the National Science foundation NSF - 2109124 and the Natural Sciences and Engineering Research Council of Canada NSERC - RGPIN-2023-03487 for their support.

B.V. also personally acknowledges Pr. Christopher Conselice for sharing observational data of NGC 1275 in the optical, as well as Dr. Rupal Mittal for sharing Herschel kinematic data of NGC 1275 in the infrared.

\software{python (\citealt{van2009python}), astropy (\citealt{robitaille_astropy_2013}, \citealt{price-whelan_astropy_2018}), numpy (\citealt{harris_array_2020}), scipy (\citealt{virtanen_scipy_2020}), matplotlib (\citealt{hunter_matplotlib_2007}), \texttt{LUCI} (\citealt{rhea_luci_2021})}

%% To help institutions obtain information on the effectiveness of their 
%% telescopes the AAS Journals has created a group of keywords for telescope 
%% facilities.
%
%% Following the acknowledgments section, use the following syntax and the
%% \facility{} or \facilities{} macros to list the keywords of facilities used 
%% in the research for the paper.  Each keyword is check against the master 
%% list during copy editing.  Individual instruments can be provided in 
%% parentheses, after the keyword, but they are not verified.

%% Similar to \facility{}, there is the optional \software command to allow 
%% authors a place to specify which programs were used during the creation of 
%% the manuscript. Authors should list each code and include either a
%% citation or url to the code inside ()s when available.

%% Appendix material should be preceded with a single \appendix command.
%% There should be a \section command for each appendix. Mark appendix
%% subsections with the same markup you use in the main body of the paper.

%% Each Appendix (indicated with \section) will be lettered A, B, C, etc.
%% The equation counter will reset when it encounters the \appendix
%% command and will number appendix equations (A1), (A2), etc. The
%% Figure and Table counter will not reset.

\clearpage

\appendix

\section{Background Variability}

\begin{figure*}[h]
    \includegraphics[width=180mm,scale=0.5]{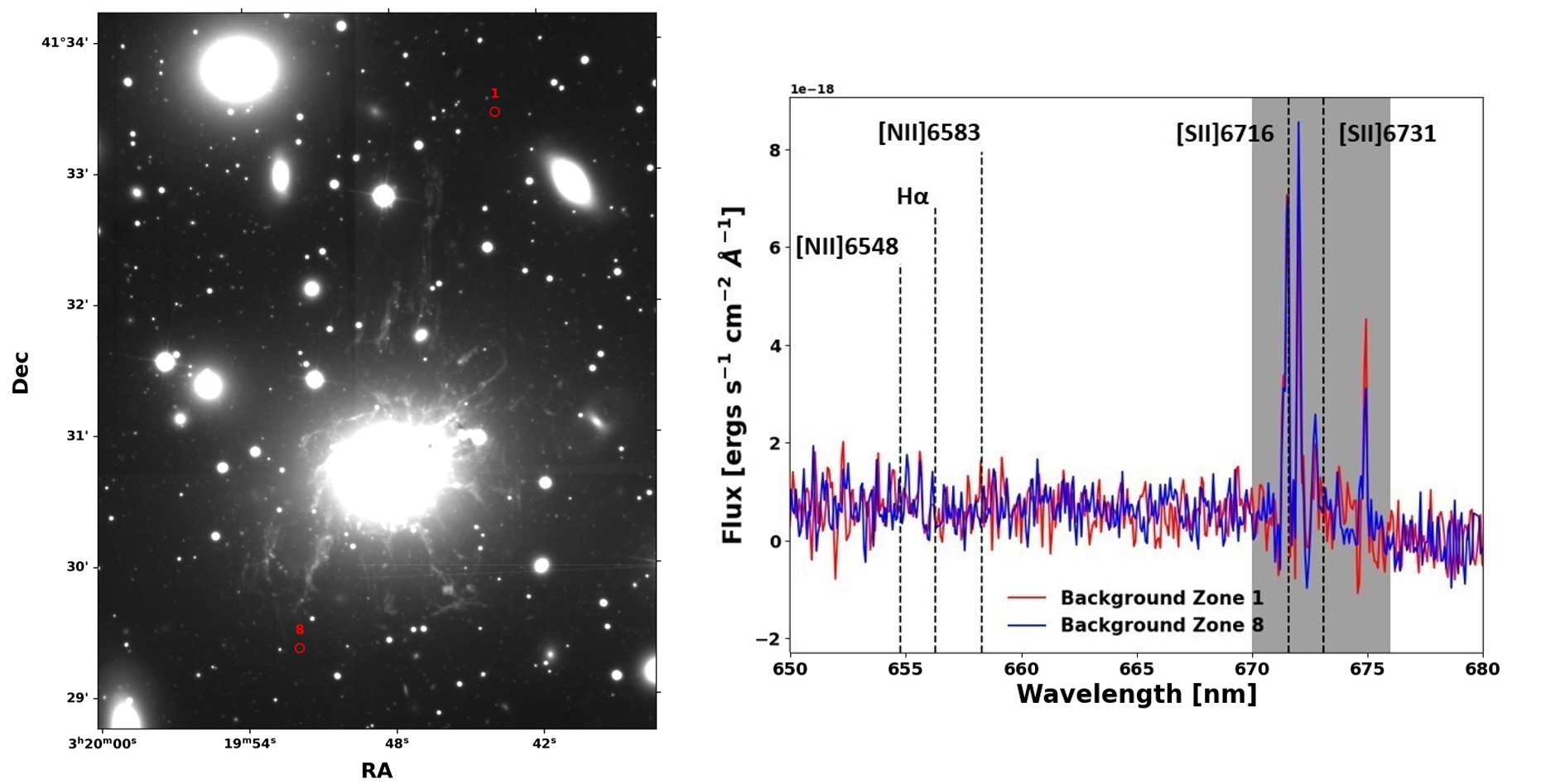}
    \caption{Comparison of the spectra extracted from two background regions denoted by the numbers 1 and 8 at high-spectral resolution. The main emission lines positions are indicated by dotted lines. The gray shaded area represents the location of the principal sky lines overlapping with the \sii\text{} emission lines doublet.}
    \label{fig:background_variability}
\end{figure*}

In this section, we display two spectra extracted from background regions (see Figure \ref{fig:background_variability}) which illustrate the clear variability in background sky emission at high-spectral resolution.

\newpage

\section{Signal-to-Noise Ratio}

\begin{figure*}[h]
    \includegraphics[width=180mm,scale=0.5]{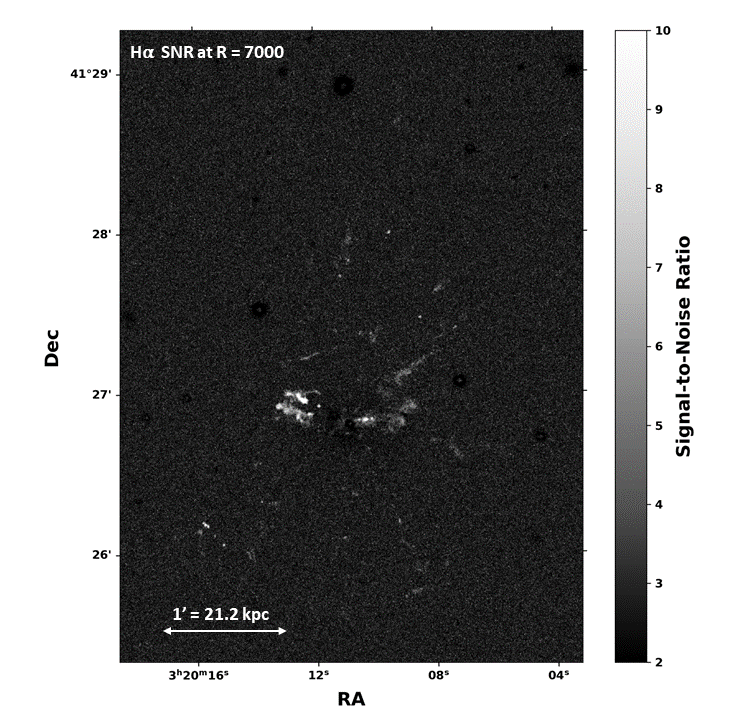}
    \caption{Signal-to-noise ratio map in linear scale of the high-spectral resolution observations of the filamentary nebula surrounding NGC 1275 with SITELLE.}
    \label{fig:signal_to_noise}
\end{figure*}

In this section, we present the signal-to-noise ratio map obtained with the \texttt{create\_snr\_map} function of \texttt{LUCI} before applying the weighted Voronoi tesselation binning. As can be seen in Figure \ref{fig:signal_to_noise}, the overall signal-to-noise ratio is low across the entire filamentary nebula and thus binning is required.

\newpage

\section{Multiple Emission Line Components Spectra}

In this section, we show a small number of spectra belonging to localized region of the filamentary nebula and displaying multiple emission line components. Their central respective coordinates are the following : RA : 3:19:48.165, DEC : +41:30:54.275 ; RA : 3:19:46.92, DEC : +41:30:51.703 ; RA : 3:19:46.623, DEC : +41:30:40.277

\begin{figure*}[h!]
    \centering
    \includegraphics[width=85mm,scale=0.5]{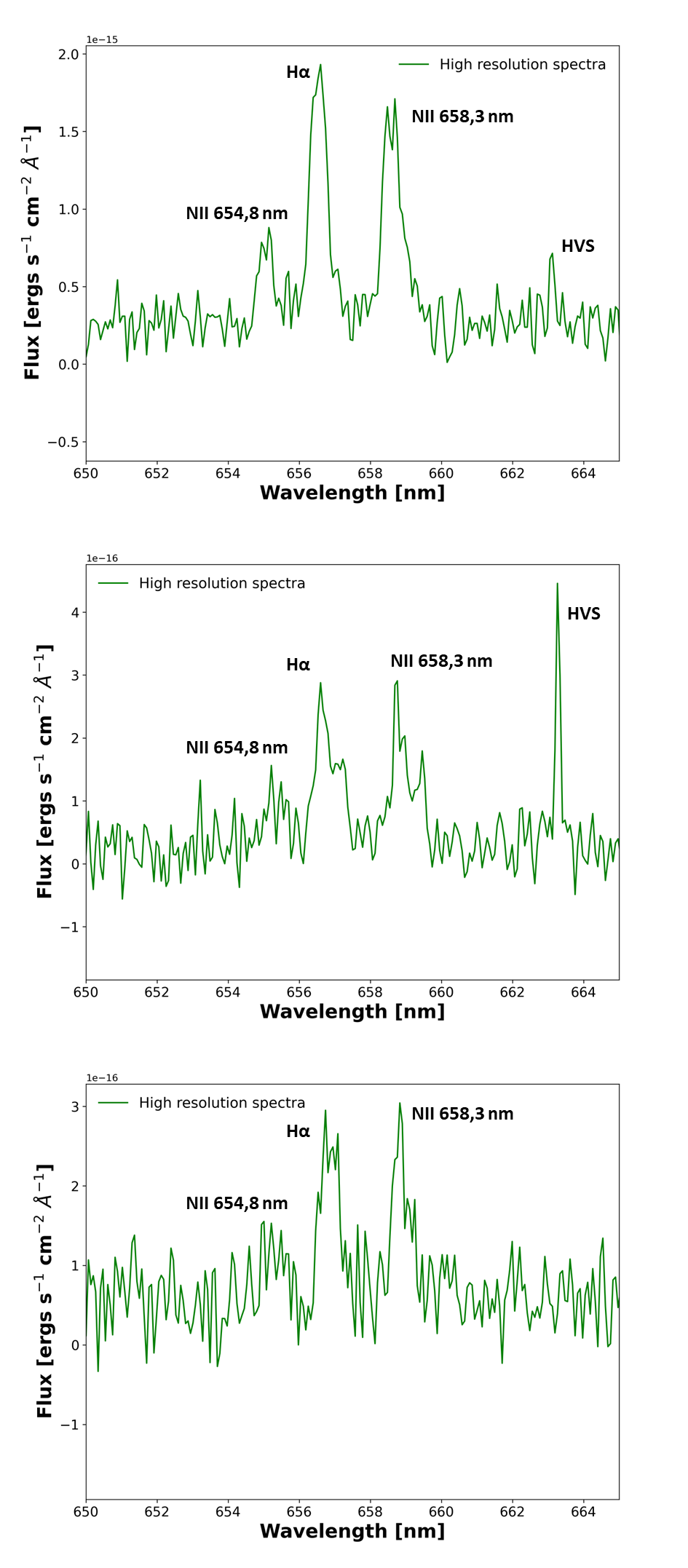}
    \caption{Spectra of localized regions of the filamentary nebula displaying multiple components emission lines.}
    \label{fig:multiple_components}
\end{figure*}

%% For this sample we use BibTeX plus aasjournals.bst to generate the
%% the bibliography. The sample63.bib file was populated from ADS. To
%% get the citations to show in the compiled file do the following:
%%
%% pdflatex sample63.tex
%% bibtext sample63
%% pdflatex sample63.tex
%% pdflatex sample63.tex

\clearpage

\bibliography{sample63}{}
\bibliographystyle{aasjournal}
% \nocite{*}
%% This command is needed to show the entire author+affiliation list when
%% the collaboration and author truncation commands are used.  It has to
%% go at the end of the manuscript.
%\allauthors

%% Include this line if you are using the \added, \replaced, \deleted
%% commands to see a summary list of all changes at the end of the article.
%\listofchanges

\end{document}